\documentclass[11pt]{article}

\usepackage{amssymb,amsthm,amscd,array}
\usepackage{graphics}
\usepackage[final]{epsfig}	

\let\ssection=\section
\renewcommand{\section}{\setcounter{equation}{0}\ssection}

\newcommand{\bbR}{\mathbb{R}}
\newcommand{\bbP}{\mathbb{P}}
\newcommand{\bbRP}{\mathbb{RP}}

\newcommand{\bbC}{\mathbb{C}}

\newcommand{\aff}{\mathrm{aff}}

\newcommand{\cD}{{\cal{D}}}
\newcommand{\cI}{{\cal{I}}}
\newcommand{\cL}{{\cal{L}}}
\newcommand{\cS}{{\cal{S}}}

\newcommand{\Diff}{\mathrm{Diff}}

\newcommand{\cF}{{\cal{F}}}

\newcommand{\Id}{\mathrm{Id}}

\newcommand{\Sl}{\mathrm{sl}}

\newcommand{\Supp}{\mathrm{Supp}}

\newcommand{\Vect}{\mathrm{Vect}}

\newcommand{\half}{\frac{1}{2}}

\newcommand{\ft}{\mathfrak{t}}
\newcommand{\fa}{\mathfrak{a}}
\newcommand{\fb}{\mathfrak{b}}

\chardef\s=110
\chardef\g=103

\begin{document}

\newtheorem{thm}{Theorem}[section]
\newtheorem{lem}[thm]{Lemma}
\newtheorem{cor}{Corollary}
\newtheorem{conj}{Conjecture}
\newtheorem{prop}[thm]{Proposition}
\newtheorem{rmk}[thm]{Remark}
\newtheorem{exe}[thm]{Example}
\newtheorem{definition}{Definition}

\def\a{\alpha}
\def\b{\beta}
\def\d{\delta}
\def\g{\gamma}
\def\om{\omega}
\def\r{\rho}
\def\s{\sigma}
\def\vfi{\varphi}
\def\vr{\varrho}
\def\l{\lambda}
\def\m{\mu}

\title{Symmetries of modules of differential operators}

\author{H. Gargoubi
\thanks{I.P.E.I.M.,
Route de Kairouan, 5019 Monastir
Tunisie, hichem.gargoubi@ipeim.rnu.tn,}
\and
P. Mathonet
\thanks{
Institut de Math\'ematique,
Grand traverse 12, B37, 4000 Li\`{e}ge, Belgique,
P.mathonet@.ulg.ac.be,}
\and
V. Ovsienko
\thanks{
CNRS, 
Institut Camille Jordan
Universit\'e Claude Bernard Lyon 1,
21 Avenue Claude Bernard,
69622 Villeurbanne Cedex,
FRANCE;
ovsienko@igd.univ-lyon1.fr
}}

\date{}

\maketitle

{\abstract{
Let $\cF_\l(S^1)$ be the space of tensor densities of degree (or weight) $\l$ on  the circle
$S^1$. The space $\cD^k_{\l,\m}(S^1)$ of $k$-th order linear  differential  operators from
$\cF_\l(S^1)$ to $\cF_\m(S^1)$ is a natural module over
$\Diff(S^1)$, the diffeomorphism group of $S^1$. We determine the algebra of symmetries of
the modules $\cD^k_{\l,\m}(S^1)$, i.e., the linear maps on $\cD^k_{\l,\m}(S^1)$
commuting with the $\Diff(S^1)$-action. We also solve the same problem in the case of
straight line $\bbR$ (instead of $S^1$) and compare the results in the compact and
non-compact cases.
}}

\thispagestyle{empty}

\section{Introduction}

We study the space of linear differential operators acting in the space of
tensor densities on $S^1$ as a module over the group $\Diff(S^1)$ of all
diffeomorphisms of $S^1$. More precisely, let $\cD^k_{\l,\m}(S^1)$ be the space of linear
$k$-th order differential operators
$$
A:\cF_\l(S^1)\to\cF_\m(S^1)
$$
where $\cF_\l(S^1)$ and $\cF_\mu(S^1)$ are the spaces of tensor densities of degree $\l$ and
$\mu$ respectively. We compute the commutant of the $\Diff(S^1)$-action on
$\cD^k_{\l,\m}(S^1)$. This commutant is an associative algebra which we denote
$\cI^k_{\l,\m}(S^1)$ and call the {\it algebra of symmetries}. 

\medskip
{\bf 1.1}
This paper is closely related to the classical subject initiated by Veblen \cite{Veb}
in his talk at IMC in 1928, namely the study of {\it invariant 
operators} also called {\it natural} operators (cf. \cite{KMS}). An operator is called 
{\it invariant} if it commutes with the action of the
group of diffeomorphisms. The main two examples are the classic de Rham differential of
differential forms
$$
d:\Omega_k(M)\to\Omega_{k+1}(M),
$$
where $M$ is a smooth manifold, and the integral
$$
\int:\Omega_n(M)\to\bbR
$$
provided $M$ is compact of dimension $n$.

Usually, one
considers differential operators acting on various spaces of tensor fields on a smooth
manifold. A famous theorem states that the de Rham differential is, actually, the only
invariant {\it differential} operator in one argument acting on the spaces of tensor fields.
This result was conjectured by Schouten and proved independently and using different
approaches by Rudakov \cite{Rud}, Kirillov \cite{KIR1} and Terng \cite{Ter}, for a complete
historical account see \cite{GLS}.

Many classification results are available now, and
it was shown that there are quite few invariant differential operators and most of them
are of a great importance. For instance, bilinear invariant differential operators on
tensor fields were classified by Grozman \cite{Gro}. The complete list of
such operators contains well-known examples, such as the Poisson, Schouten and Nijenhuis
brackets, and one exceptional bilinear third-order differential operator
\begin{equation}
\label{GroOpDef}
G:\cF_{-\frac{2}{3}}(S^1)\otimes\cF_{-\frac{2}{3}}(S^1)\to\cF_{\frac{5}{3}}(S^1).
\end{equation}

Note that differential operators invariant with respect of the diffeomorphism groups can be
interpreted in terms of the Lie algebras of vector fields. This viewpoint relates the subject
with the Gelfand-Fuchs cohomology, see
\cite{Fuc} and references therein.

\medskip
{\bf 1.2}
The main difference of our work from the classic literature is that we consider
linear operators acting on differential operators (instead of tensor fields). More
precisely, we classify the linear maps
\begin{equation}
T:\cD^k_{\l,\m}(S^1)\to\cD^k_{\l,\m}(S^1)
\label{Endom}
\end{equation}
commuting with the $\Diff(S^1)$-action.
The module of differential operators $\cD^k_{\l,\m}(S^1)$ is
not isomorphic to any module of tensor fields (but rather rasembles $\mathrm{gl}(\cF_\l)$ or
$\cF_\l^*\otimes\cF_\mu$). The problem of classification of
$\Diff(S^1)$-invariant operators on $\cD^k_{\l,\m}(S^1)$ is, therefore, different from
Veblen's problem, although similar.

A well-known example of a map (\ref{Endom}) is the \textit{conjugation}
of differential operators. This map associates to an operator $A$ the \textit{adjoint}
operator $A^*$. If $A\in\cD^k_{\l,\m}(S^1)$, then $A^*\in\cD^k_{1-\m,1-\l}(S^1)$, so that 
this map defines a symmetry if and only if
$\l+\m=1$.

Let us emphasize that, unlike Rudakov-Kirillov-Terng, we consider
not only differential (or local) symmetries of
$\cD^k_{\l,\m}(S^1)$ but also \textit{non-local} ones. For instance, we find a version of
trace which is an analog of the Adler trace (see \cite{Adl}).

\medskip
{\bf 1.3}
The main purpose of this paper is to show that some modules $\cD^k_{\l,\m}(S^1)$ are
particular and very interesting.
It turns out that the algebra of symmetries $\cI^k_{\l,\m}(S^1)$ can be quite rich
depending on the values of $\l$ and $\m$ as well as on $k$. This
algebra is an important characteristic of the corresponding module which embraces those given
in
\cite{GO,GAR,GO1}.

The first example which is particular (for every $k$) is the module
$\cD^k_{0,1}(S^1)$ of operators from the space of functions to the space of 1-forms. The
algebra of symmetries in this case is always of maximal dimension.

Another interesting module is
$\cD^2_{-\half,\frac{3}{2}}(S^1)$. It is related to the famous Virasoro algebra and
also to the projective differential geometry, see \cite{KIR,OT}. This module
appears as a particular case in our classification. 

Further intriguing
examples of modules of differential operators are
$\cD^3_{-\frac{2}{3},\frac{5}{3}}(S^1)$
and $\cD^4_{-\frac{2}{3},\frac{5}{3}}(S^1)$.
These modules are related to the Grozman operator (\ref{GroOpDef}). The algebraic and
geometric meaning of these modules is not known.

\medskip
{\bf 1.4}
We will also consider symmetries of differential operators acting in the space of
$\l$-densities over $\bbR$ and compare this case with the case of $S^1$. The classification
of the invariant differential operators (\ref{Endom}) remains the same as that on $S^1$,
except that there are no non-local symmetries.

Let us also mention that symmetries of the modules of differential operators in
the multi-dimensional case have been classified in \cite{MAT}. In this case the
algebra of symmetries is smaller.

\medskip
{\bf 1.5}
This paper is organized as follows. In Section \ref{MainDefSec} we introduce the
$\Diff(S^1)$-modules of differential operators and their symbols. In Section
\ref{MainResSec} we formulate the classification theorems. In Section
\ref{ConstSymm} we give an explicit construction of all $\Diff(S^1)$-invariant linear maps
on the modules $\cD^k_{\l,\m}(S^1)$ in all possible cases. These operators are our main
characters; some of them are known and some seem to be new. In Section \ref{ComputingSec}
we calculate the associative algebras of invariant operators and identify them with
some associative algebras of matrices. This gives a complete description of the symmetry
algebras $\cI^k_{\l,\m}(S^1)$. Finally, in Section \ref{EndProof} we prove that there are no
other invariant operators on $\cD^k_{\l,\m}(S^1)$ than the operators we introduce. This
completes the proof of the main theorems.

\section{The main definitions}\label{MainDefSec}

In this section we define the space of differential operators on $S^1$ acting on
the space of densities and the corresponding space of symbols. We pay particular attention to
the action of the group of diffeomorphisms $\Diff(S^1)$ and of the Lie algebra of vector
fields $\Vect(S^1)$ on these spaces.

\subsection{Densities on $S^1$}

Denote by $\cF_\l(S^1)$, or $\cF_\l$ for short, the space of $\l$-densities on $S^1$
$$
\varphi=\phi(x)(dx)^\l,
$$
where $\l\in\bbC$ is the \textit{degree} (or \textit{weight}), $x$ is a local
coordinate on $S^1$ and
$\phi(x)\in{}C^\infty(S^1)$. As a vector space, $\cF_\l$ is isomorphic to $C^\infty(S^1)$.

The group $\Diff(S^1)$ naturally acts on $\cF_\l$. If $f\in\Diff(S^1)$, then
$$
\rho^\l_{f^{-1}}:\phi(x)(dx)^\l
\mapsto
\left(f'\right)^\l
\,\phi(f(x))(dx)^\l\,.
$$
The $\Diff(S^1)$-modules $\cF_\l$ and $\cF_\m$ are not isomorphic unless $\l=\m$ (cf.
\cite{Fuc}). 

\begin{exe}
{\rm
The space $\cF_0$ is isomorphic to $C^\infty(S^1)$, as a $\Diff(S^1)$-module;
the space $\cF_1$ is nothing but the space of 1-forms (volume forms); the space $\cF_{-1}$ is
the space of vector fields on $S^1$. 
}
\end{exe}

The space $\cF_\l$ can also be viewed as the space of functions on the
cotangent bundle $T^*S^1\setminus{}S^1$ (with zero-section removed) homogeneous of degree
$-\l$. In standard (Darboux) coordinates
$(x,\xi)$ on $T^*S^1$ one writes:
\begin{equation}
\label{IdenDenFun}
\phi(x)(dx)^\l
\leftrightarrow
\phi(x)\,\xi^{-\l}.
\end{equation}
This identification commutes with the
$\Diff(S^1)$-action.

\subsection{Invariant pairing}

There is a pairing $\cF_\l\otimes\cF_{1-\l}\to\bbR$ given by
$$
\left\langle\phi(x)(dx)^\l,\psi(x)(dx)^{1-\l}\right\rangle=
\int_{S^1}\phi(x)\psi(x)dx
$$
which is $\Diff(S^1)$-invariant.
For instance, the space $\cF_\half$ is equipped with a scalar product; this is a natural
pre-Hilbert space that is popular in geometric quantizaton.

\subsection{Differential operators on densities}

Consider the space of
linear differential operators
$$
A:\cF_\l\to\cF_\m
$$
with arbitrary $\l,\m\in\bbC$.
This space will be denoted by $\cD_{\l,\m}(S^1)$. The subspace of differential
operators of order $\leq k$ will be denoted by $\cD^k_{\l,\m}(S^1)$.

Fix a (local) coordinate $x$, a differential operator $A\in\cD^k_{\l,\m}(S^1)$ is of the form
\begin{equation}
\label{DiOpA}
A=a_k(x)\frac{d^k}{dx^k}+
a_{k-1}(x)\frac{d^{k-1}}{dx^{k-1}}+\cdots+a_0(x),
\end{equation}
where $a_i(x)$ are smooth functions.
More precisely,
$$
A(\varphi)
=\left(
a_k(x)\frac{d^k\phi(x)}{dx^k}+
a_{k-1}(x)\frac{d^{k-1}\phi(x)}{dx^{k-1}}+\cdots+a_0(x)\phi(x)
\right)
(dx)\mu,
$$
where $\varphi=\phi(x)(dx)^\l$.

\begin{exe}
{\rm
The space $\cD^0_{\l,\m}(S^1)$ is nothing but $\cF_{\mu-\l}$. Any zeroth-order differential
operator is the operator of multiplication by a $(\mu-\l)$-density:
$$
a(x)(dx)^{\mu-\l}:\phi(x)(dx)^\l
\mapsto
a(x)\,\phi(x)(dx)^\m
$$
}
\end{exe}

\subsection{$\Diff(S^1)$- and $\Vect(S^1)$-module structure}

The space $\cD_{\l,\m}(S^1)$ is a $\Diff(S^1)$-module with respect to the action
$$
\rho^{\l,\m}_f(A)=
\rho^{\m}_f\circ{}A\circ\rho^{\l}_{f^{-1}},
$$
where $f\in\Diff(S^1)$.

We will also consider the Lie algebra of vector fields $\Vect(S^1)$
and the natural $\Vect(S^1)$-action on $\cD^k_{\l,\m}(S^1)$.
A vector field $X=X(x)\,\frac{d}{dx}$ acts on the space of tensor
densities $\cF_\l$ by Lie derivative
$$
L^\l_X(\varphi)=
\left(X(x)\,\phi'(x)+
\l\,
X'(x)\,\phi(x)\right)\,(dx)^\l.
$$
The action of $\Vect(S^1)$ on the space of differential operators is given by the commutator
\begin{equation}
\cL^{\l,\m}_X(A)=
L^\m_X\circ{}A-A\circ{}L^\l_X.
\label{comm}
\end{equation}
Note that the above formul{\ae} are independent of the choice of the local coordinate $x$.

\subsubsection{Example: the module $\cD^1_{\l,\m}(S^1)$}

The space $\cD^1_{\l,\m}(S^1)$ is split into a direct sum
\begin{equation}
\label{SplOn}
\cD^1_{\l,\m}(S^1)
\cong\cF_{\mu-\l-1}\oplus\cF_{\mu-\l}
\end{equation}
as a $\Diff(S^1)$- (and $\Vect(S^1)$-) module.

Indeed, every first-order differential operator $A\in\cD^1_{\l,\m}(S^1)$
$$
A\left(\phi(x)(dx)^\l\right)
=
\left(a_1(x)\,\phi'(x)+a_0(x)\,\phi(x)\right)(dx)^\mu
$$
can be rewritten in the form
$$
A\left(\phi(dx)^\l\right)
=
\left(a_1\,\phi'+\l\,a_1'\,\phi+
(a_0-\l\,a_1')\,\phi\right)(dx)^\mu,
$$
and, finally, one obtaines an intrinsic expression
$$
A(\varphi)=\left(L_{a_1}\varphi+(a_0-\l\,a_1')\varphi\right)
(dx)^{\mu-\l},
$$
where $a_1=a_1(x)\frac{d}{dx}$ is understood a vector field and $a_0(x)-\l\,a_1'(x)$ as a
function.

Furthermore, using identification (\ref{IdenDenFun}), one can write a more elegant formula
for a first-order operator:
$$
A=\xi^{\mu-\l}\,L_{a_1}+\mathrm{div}{a_1}\frac{\partial}{\partial\xi}+a_0.
$$

Note that there are no intrinsic formul{\ae} similar to the above ones in the case of modules
$\cD^k_{\l,\m}(S^1)$ with $k\geq2$, and, in general, there are no splittings similar to
(\ref{SplOn}).  The geometric meaning of the modules $\cD^k_{\l,\m}(S^1)$ was
discussed in \cite{GO,BO,GAR,GO1}.

\subsection{Space of symbols of differential operators}

The filtration
$$
\cD^0_{\l,\m}(S^1)\subset\cD^1_{\l,\m}(S^1)\subset
\cdots\subset\cD^k_{\l,\m}(S^1)\subset\cdots
$$
is preserved by the $\Diff(S^1)$-action.  
The graded $\Diff(S^1)$-module
$
\cS_{\l,\m}(S^1)={\rm gr}\,(\cD_{\l,\m}(S^1))
$
is called the module of \textit{symbols} of differential operators.

The quotient module $\cD^k_{\l,\m}(S^1)/\cD^{k-1}_{\l,\m}(S^1)$ is
isomorphic to the module of tensor densities $\cF_{\m-\l-k}(S^1)$; the isomorphism is
provided by the principal symbol. As a $\Diff(S^1)$-module, the space of
symbols depends, therefore, only on the difference
$$
\d=\m-\l,
$$
so that $\cS_{\l,\m}(S^1)$ can be denoted as $\cS_{\d}(S^1)$, and finally we have
$$
\cS_{\d}(S^1)=\bigoplus_{i=0}^\infty\cF_{\d-i}
$$
as  $\Diff(S^1)$-modules.

The space of symbols $\cS_{\d}(S^1)$ can also be viewed as the space of functions on
$T^*S^1\setminus{}S^1$. Namely, any $k$-th order symbol $P\in\cS_{\d}(S^1)$ can be written
in the form
\begin{equation}
\label{SymDiOpA}
P=a_k(x)\,\xi^{k-\d}+
a_{k-1}(x)\,\xi^{k-\d-1}+\cdots+a_0(x)\,\xi^{-\d}.
\end{equation}
The natural lift of the action of $\Diff(S^1)$ to $T^*S^1$ equips the space of functions
(\ref{SymDiOpA}) with a structure of $\Diff(S^1)$-module; this action coincides with
the $\Diff(S^1)$-action on the space $\cS_{\d}(S^1)$.

\begin{rmk}
{\rm
The spaces $\cD_{\l,\m}(S^1)$ and $\cS_{\d}(S^1)$ are not isomorphic as
$\Diff(S^1)$-modules. There are cohomology classes which are obstructions for existence of
such an isomorphism, see \cite{LO1,BO,GO}.
}
\end{rmk}

\section{The main results}\label{MainResSec}

In this section we formulate the main results of this paper and give a complete
description of the algebras of symmetries $\cI^k_{\l,\m}(S^1)$.
We defer the proofs to Sections \ref{ConstSymm}-\ref{EndProof}.

We will say that the algebra $\cI^k_{\l,\m}(S^1)$ is \textit{trivial} if it is 
generated by the identity map~$\Id$, and therefore is isomorphic to $\bbR$. Of
course, we are interested in the cases when this algebra is non-trivial.

\subsection{Introducing four algebras of matrices}\label{martix}

We will need the following associative algebras.

\begin{enumerate}

\item 
The commutative algebra $\bbR\oplus\cdots\oplus\bbR$ will be denoted by~$\bbR^n$. Of course,
this algebra can be represented by diagonal $n\times{}n$-matrices.

\medskip
\item
The algebra of (lower) triangular $(n\times\!{}n)$-matrices will be denoted by
$\mathfrak{t}_n$.

\medskip
\item
The commutative algebra $\fa$ of $(2\times\!2)$-matrices of the form
$$
\left(\begin{array}{cc}a & 0\\ b & a\end{array}\right).
$$

\medskip
\item
The algebra $\fb$ of $(4\times\!4)$-matrices of the form
$$
\left(\begin{array}{cccc}
a & 0 & 0 & d\\
0 & a & 0 & 0\\
0 & c & b &0\\
0&0&0&b
\end{array}\right)
$$

\end{enumerate}

It turns out that the algebras of symmetries $\cI^k_{\l,\m}(S^1)$ are always
direct sums of the above matrix algebras.

In Appendix we will introduce natural generators of the algebras $\fa$, $\fb$ and
$\bbR^n$.

\subsection{Stability: the case $k\geq5$}

We start our list of classification theorems with the ``stable'' case. Namely, 
if $k\geq5$, then the algebras $\cI^k_{\l,\m}(S^1)$ do not depend on $k$.

\begin{thm}
\label{Dim1Thm1}
For $k\geq5$, the algebra $\cI^k_{\l,\m}(S^1)$ is trivial
for all $(\l,\m)$, except

\begin{enumerate}

\item 
$\cI^k_{\l,\m}(S^1)\cong\bbR^2$,
\hskip1cm for 
$
\left\{
\begin{array}{ll}
\l+\m=1, &\l\neq0,\\
\l=0, &\m\not=1,\,0\\
 \m=1, &\l\not=1,\,0
\end{array}
\right.$

\medskip
\item
$\cI^k_{0,0}(S^1)\cong\cI^k_{1,1}(S^1)\cong\bbR^3$;

\medskip
\item
$\cI^k_{0,1}(S^1)\cong\fb\oplus\bbR^2$.
\end{enumerate}
\end{thm}

\begin{figure}[ht]
\centerline{\epsfbox{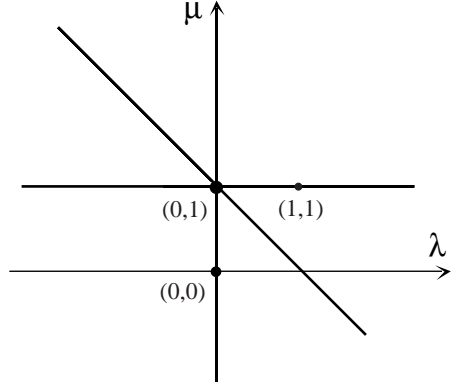}}
\caption{Exceptional modules of higher order operators}
\label{FigclassOdin}
\end{figure}

\noindent
The exceptional modules $\cD^k_{\l,\m}(S^1)$ with $k\geq5$ are represented
in Figure \ref{FigclassOdin}.

We will provide in the sequel a list of generators of the symmetry algebra for each
non-trivial case, we will also give its explicit identification with the corresponding
algebra of matrices.

\subsection{Modules of differential operators of order 4}

The result for the modules $\cD^k_{\l,\m}(S^1)$ with $k\leq4$ is different from the above
one. (It is interesting to compare this property of differential operators with that
in the case of algebraic equations.)

Consider the modules of operators of order $k=4$. The complete classification of
symmetries in this case is given by the following result.

\begin{thm}
\label{Dim1Thm2}
The algebra $\cI^4_{\l,\m}(S^1)$ is trivial
for all $(\l,\m)$ except

\begin{enumerate}

\item 
 $\cI^4_{\l,\m}(S^1)\cong\bbR^2,$ 
\hskip1cm
 for 
$
\left\{
\begin{array}{ll}
\l+\m=1, &\l\neq0,\,-\frac{2}{3}\\[4pt]
\l=0, &\m\not=3,\,\frac{5}{4},\,1,\,0\\[4pt]
 \m=1, &\l\not=1,\,0,\,-\frac{1}{4},\,-2
\end{array}
\right.$

\medskip
\item 
$\cI^4_{\l,\m}(S^1)\cong\bbR^3,$
\hskip1cm
 for 
$(\l,\m)=(1,1),\,(0,\frac{5}{4}),\,(0,0),\,(-\frac{1}{4},1),\,(-\frac{2}{3},\frac{5}{3})$;

\medskip
\item 
$\cI^4_{0,3}(S^1)\cong\cI^4_{-2,1}(S^1)\cong
\fa\oplus\bbR$;

\medskip
\item 
$\cI^4_{0,1}(S^1)\cong\fb\oplus\bbR^2$.

\end{enumerate}
\end{thm}

\begin{figure}[ht]
\centerline{\epsfbox{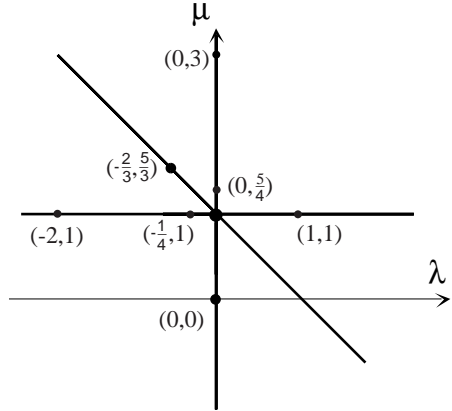}}
\caption{Exceptional modules of 4-th order operators}
\label{FigclassFour}
\end{figure}

\noindent
The exceptional modules $\cD^4_{\l,\m}(S^1)$ are represented in Figure
\ref{FigclassFour}.

\begin{rmk}
{\rm
We will show in  Section \ref{explicit} that the module
$\cD^4_{-\frac{2}{3},\frac{5}{3}}(S^1)$ is, indeed, a very special one. This exceptional 
module is related to the Grozman operator (\ref{GroOpDef}).
}
\end{rmk}

\subsection{Modules of differential operators of order 3}

Symmetries of the modules of third-order operators are particularly rich.

\begin{thm}
\label{Dim1Thm3}
The algebra $\cI^3_{\l,\m}(S^1)$ is trivial
for all $(\l,\m)$ except
\begin{enumerate}

\item 
$\cI^3_{\l,\m}(S^1)\cong\bbR^2$,
\hskip1cm for 
$
\left\{
\begin{array}{ll}
\l+\m=1, &\l\neq0,-\frac{1}{2},-\frac{2}{3}\\[4pt]
(3\l+1)(3\m-4)=-1, &\l\not=0,-\frac{2}{3}
\end{array}
\right.$

\medskip
\item 
$\cI^3_{\l,\m}(S^1)\cong\fa$,
\hskip1,3cm for 
$\m-\l=2$, $\l\neq0,-\frac{1}{2},-1$;

\medskip
\item 
$\cI^3_{\l,\m}(S^1)\cong\bbR^3$, 
\hskip1cm for 
$
\left\{
\begin{array}{ll}
\l=0, &\m\not=3,\,2,\,1\\
\m=1, &\l\not=0,-1,-2
\end{array}
\right.$

\medskip
\item 
$\cI^3_{\l,\m}(S^1)\cong\fa\oplus\bbR$,
\hskip1cm
for $(\l,\m)=(0,3),\,(0,2),\,(-1,1),\,(-2,1)$;

\medskip
\item 
$\cI^3_{-\frac{1}{2},\frac{3}{2}}(S^1)\cong\ft_2$;

\medskip
\item 
$\cI^3_{-\frac{2}{3},\frac{5}{3}}(S^1)\cong\bbR^3$;

\medskip
\item 
$\cI^3_{0,1}(S^1)\cong\fb\oplus\bbR^2$.

\end{enumerate}
\end{thm}

\begin{figure}[ht]
\centerline{\epsfbox{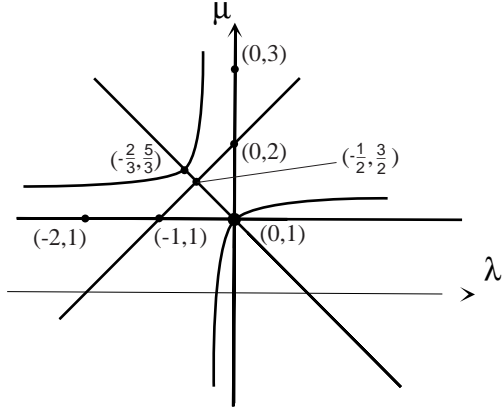}}
\caption{Exceptional modules of third-order operators}
\label{FigclassThree}
\end{figure}

\noindent
The exceptional modules $\cD^3_{\l,\m}(S^1)$ are represented in
Figure \ref{FigclassThree}.

\subsection{Modules of second-order differential operators}

Second-order differential operators are definitely among the most popular
objects of mathematics. Their invariants with respect to the action of $\Diff(S^1)$,
such as monodromy or the rotation number, were thoroughly studied. Some modules of
second-order differential operators on
$S^1$ have geometric meaning, they have been related to various algebraic
structures such as the Virasoro algebra and integrable systems.

The result in the second-order case is as follows.

\begin{thm}
\label{Dim1Thm4}
The algebra $\cI^2_{\l,\m}(S^1)$ is isomorphic to 
$\bbR^2$ for all $(\l,\m)$, except

\begin{enumerate}

\item 
$\cI^2_{\l,\m}(S^1)\cong\fa$,
\hskip1,3cm for
$\left\{
\begin{array}{ll}
\m-\l=1, & \l\neq0\\[4pt]
\m-\l=2, & \l\neq0,\,-\half,\,-1
\end{array}
\right.$

\medskip
\item 
$\cI^2_{\l,\m}(S^1)\cong\bbR^3$,
\hskip1cm for 
$
\left\{
\begin{array}{ll}
\l=0, &\m\not=2,\,1\\
\m=1, &\l\not=0,\,-1
\end{array}
\right.$

\medskip
\item 
$\cI^2_{-\half,\frac{3}{2}}(S^1)
\cong\ft_2$;

\medskip
\item 
$\cI^2_{0,2}(S^1)\cong\cI^2_{-1,1}(S^1)
\cong\fa\oplus\bbR$;

\medskip
\item 
$\cI^2_{0,1}(S^1)\cong\fb\oplus\bbR$.

\end{enumerate}
\end{thm}

\begin{figure}[ht]
\centerline{\epsfbox{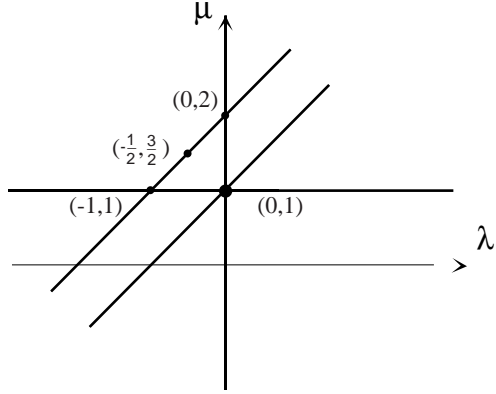}}
\caption{Exceptional modules of second-order operators}
\label{FigclassTwo}
\end{figure}

\noindent
The exceptional modules of second-order operators are
represented in Figure \ref{FigclassTwo}.

\subsection{Modules of first-order differential operators}

Let us finish this section with the result in the first-order case. The result is less
interesting, the only particular module is $\cD^1_{0,1}(S^1)$.

\begin{thm} 
\label{Dim1Thm5}
The algebra of $\cI^1_{\l,\m}(S^1)$ is isomorphic to
$\bbR^2$ for all $\l,\m$, except in the following cases:

\begin{enumerate}

\item 
$\cI^1_{\l,\m}(S^1)\cong\fa$, for $\m-\l=1$, $\l\not=0$;

\medskip

\item 
$\cI^1_{0,1}(S^1)\cong\fb$.

\end{enumerate}
\end{thm}

For the sake of completeness, let us also mention the zeroth-order case:
the algebra of symmetries $\cI^0_{\l,\m}(S^1)$ is trivial. Indeed, the module
$\cD^0_{\l,\m}(S^1)$ is isomorphic to $\cF_{\m-\l}$.

\subsection{Non-compact case: differential operators on $\bbR$}

The algebras of symmetries of the modules of differential operators in the spaces of
densities over $\bbR$ can be different from that on $S^1$. This occurs in the ``most
particular'' case $(\l,\m)=(0,1)$. 

\begin{thm}
(i)
If $(\l,\m)\not=(0,1)$, then the algebra $\cI^k_{\l,\m}(\bbR)$ coincides with
$\cI^k_{\l,\m}(S^1)$. 

(ii)
In the exceptional case $(\l,\m)=(0,1)$ one has:
\begin{enumerate}

\item
If $k\geq3$, then $\cI^k_{0,1}(\bbR)\cong\ft_2\oplus\bbR^2$;

\medskip

\item 
If $k=2$, then $\cI^k_{0,1}(\bbR)\cong\ft_2\oplus\bbR$;

\item 
If $k=1$, then $\cI^k_{0,1}(\bbR)\cong\ft_2$.

\end{enumerate}
\end{thm}

\section{Construction of symmetries}\label{ConstSymm}

In this section we give an explicit construction of the generators of algebras
$\cI^k_{\l,\m}(S^1)$ for every case where this algebra is non-trivial. We will prove in
Section \ref{EndProof} that our list of invariant differential operators is complete.

\subsection{The conjugation}

The best known invariant map between the spaces of differential operators is
the \textit{conjugation}. It is a linear map
$$
C:\cD^k_{\lambda,\m}(S^1)\to
\cD^k_{1-\m,1-\lambda}(S^1)
$$
that associates to each operator $A$ its adjoint $A^*$ defined by
$$
\int_{S^1} A^*(\varphi)\psi=
\int_{S^1} \varphi A(\psi)
$$
for every $\varphi\in{\cal F}_{1-\m}$ 
and $\psi\in{\cal F}_{\lambda}$.
It follows that the modules $\cD^k_{\l,\m}(S^1)$ with $\l$ and $\m$ satisfying the
condition
$$
\l+\m=1
$$
have non-trivial symmetries.

The conjugation map $C$ is an {\it involution}; the straight line $\l+\m=1$ will play a
role of symmetry axis in the plane parameterized by $(\l,\m)$.

For an arbitrary local parameter $x$ on $S^1$, the conjugation map is 
given by the well-known formula
\begin{equation}
C:
\sum_{i=0}^ka_i(x)\frac{d^i}{dx^i}
\mapsto
\sum_{i=0}^k
(-1)^i\,
\left(\frac{d}{d{}x}\right)^i\circ
a_i(x)
\label{CA}
\end{equation}
that easily follows from the definition.

\begin{rmk}
{\rm
The expression (\ref{CA}) is independent from the choice of the parameter
$x$. Indeed, any change of local coordinates is given by a diffeomorphism of
$S^1$. Note that this fundamental property of coordinate independence is just a
different way to express the $\Diff(S^1)$-equivariance.
}
\end{rmk}

\subsection{The cases $\l=0$ and $\m=1$}\label{l=0m=1}

We will define a $\Diff(S^1)$-invariant operator
$$
P_0:\cD^k_{0,\m}(S^1)\to\cD^k_{0,\m}(S^1).
$$
Let us first consider a $\Diff(S^1)$-invariant projection
$
P_0:\cD^k_{0,\m}(S^1)\to{\cal F}_{\m}
$
defined by: 
$
A\mapsto A(1),
$
where $1\in\cF_0\cong{}C^{\infty}(M)$ is a constant function on $S^1$.
In other words, 
\begin{equation}
P_0\left(
\sum_{i=0}^ka_i(x)\frac{d^i}{dx^i}
\right)
=a_0(x)\left(dx\right)^\m.
\label{pro}
\end{equation}
Since $\cF_\m\subset\cD^k_{0,\m}(S^1)$, one obtains a non-trivial element of the
algebra $\cI^k_{0,\m}(S^1)$.

Thanks to the conjugation map (\ref{CA}), one also has a non-trivial symmetry
$P_0^*=C\circ{}P_0\circ{}C$
$$
P_0^*:\cD^k_{\l,1}(S^1)\to\cD^k_{\l,1}(S^1).
$$
The explicit formula follows from (\ref{pro}) and (\ref{CA}):
\begin{equation}
\label{leP0'expl}
P_0^*\left(
\sum_{i=0}^ka_i(x)\frac{d^i}{dx^i}
\right)
=
\sum_{i=0}^k(-1)^i\,a_i(x)^{(i)}.
\end{equation}
The right hand side is understood as a (scalar) differential operator from $\cF_\l$ to
$\cF_1$.

\subsection{Two additional elements of $\cI^k_{0,1}(S^1)$}

In the most particular case $(\l,\m)=(0,1)$, there are two more elements of the symmetry
algebra.

\begin{itemize}
\item
There is a \textit{non-local} element of $\cI^1_{0,1}(S^1)$.
It is given by the expression
\begin{equation}
\label{leL}
L\left(
\sum_{i=0}^ka_i(x)\frac{d^i}{dx^i}
\right)=
\left(\int_{S^1}a_0(x)\,dx\right)
\circ{}d
\end{equation}
where $d$ is the de Rham differential. 
Indeed, the projection (\ref{pro}) gives a 1-form on $S^1$ so that the integral above is
well-defined; since $d\in\cD^1_{0,1}(S^1)$, we can understand $L$ as a linear map
$L:\cD^k_{0,1}(S^1)\to\cD^k_{0,1}(S^1)$ and therefore a symmetry.

\begin{rmk}
{\rm
Equation (\ref{leL}) is an analog of the well-known Adler trace \cite{Adl},
although the latter is defined on the space of pseudodifferential operators from $\cF_0$ to
$\cF_0$.
}
\end{rmk}

\item
There is one more element of the algebra 
$\cI^1_{0,1}(S^1)$ given by the formula
\begin{equation}
\label{leP1}
P_1\left(
\sum_{i=0}^ka_i(x)\frac{d^i}{dx^i}
\right)=
\left(
\sum_{i=1}^k(-1)^{i-1}\,a_i(x)^{(i-1)}
\right)
\circ{}d.
\end{equation}
It is easy to check directly that $P_1$ is $\Diff(S^1)$-invariant, but its intrinsic form can
also be written.
In the one-dimensional case, the de Rham differential $d$ is an element of
$\cD^1_{0,1}(S^1)$. One has a $\Diff(S^1)$-invariant operator
$$
\d:\cD^k_{1,\m}(S^1)\to\cD^{k+1}_{0,\m}(S^1)
$$
given by right composition with the de Rham differential: 
$
\d:A\mapsto{}A\circ{}d.
$
This map is a bijection between $\cD^k_{1,\m}(S^1)$ and
$\mathrm{Ker}\,P_0\subset\cD^{k+1}_{0,\m}(S^1)$.
One has: 
$$
P_1=\d\circ{}P_0\circ{}C\circ\d^{-1}\circ(\Id-P_0).
$$
\end{itemize}

\subsection{Additional elements of
$\cI^k_{0,0}(S^1)$ and $\cI_{1,1}^k(S^1)$}\label{explicitBBis}

The algebra $\cI^k_{0,0}(S^1)$ is generated by the operator $P_0$ given by (\ref{pro}) and
\begin{equation}
\label{Tdelta}
S = -C\circ\d^{-1}\circ(\Id - P_{0})\circ C\circ \d\circ C.
\end{equation}
One can check that the explicit formula for this operator is as
follows:
$$
S\left(
\sum_{i=0}^ka_i(x)\frac{d^i}{dx^i}
\right)=
\sum_{i=0}^{k-1}
(-1)^i\left(\frac{d}{dx}\right)^i\circ
\left(
a_i(x)+a_{i+1}'(x)
\right)
+(-1)^k\left(\frac{d}{dx}\right)^k\circ
a_k(x).
$$

The algebra $\cI^k_{1,1}(S^1)$ is generated by $P_0^*$ given by (\ref{leP0'expl}) and the
operator $S^*=C\circ{}S\circ{}C$. 

\subsection{Symmetries and bilinear operators on tensor densities}
\label{SBOTDSec}

We now give a general way to construct linear $\Diff(S^1)$-invariant differential operators
on $\cD^k_{\l,\m}(S^1)$.
Assume there are two $\Diff(S^1)$-invariant differential operators:
\begin{itemize}
\item
a \textit{bilinear} differential operator
$
J:{\cal F}_{\nu}\otimes{\cal F}_{\l}\to{\cal F}_{\m};
$
\item
a linear projection
$
\pi:\cD^k_{\l,\m}(S^1)\to{\cal F}_{\nu}.
$
\end{itemize}
We define a linear map 
$
J\circ\pi:\cD^k_{\l,\m}(S^1)\to\cD^k_{\l,\m}(S^1)
$
as follows
$$
\left(
J\circ\pi\right)(A)(\cdot)=
J\left(\pi(A),\,\cdot\right).
$$
This map is obviously $\Diff(S^1)$-invariant.

This is the way invariant differential operators on the modules
$\cD^k_{\l,\m}(S^1)$ are related to invariant bilinear differential operators on 
densities. We give here the complete list of bilinear operators on densities
and the complete list of linear projections. We will then specify the generators of the
algebras $\cI^k_{\l,\m}(S^1)$ that can be obtained by the above construction.

\subsubsection{Bilinear invariant differential operators on tensor densities}\label{bil}

The classification of invariant bilinear differential operators on tensor fields is due
to P. Grozman \cite{Gro}. His list is particularly interesting
in the one-dimensional case (see also \cite{FF1}). 

Let us recall here the complete list.

\begin{enumerate}

\item 
Every zeroth-order operator
$\cF_{\nu}\otimes\cF_{\l}\to\cF_{\nu+\l}$ is of the form:
$$
\phi(x)(dx)^\l\otimes\psi(x)(dx)^\mu\mapsto
c\,\phi(x)\psi(x)(dx)^{\l+\m},
$$
where $c\in\bbC$. From now on we omit a scalar multiple $c$.

\medskip

\item 
Every first order operator
$\cF_{\nu}\otimes\cF_{\l}\to\cF_{\nu+\l+1}$
is as follows
\begin{equation}
\left\{\phi(x)(dx)^\nu,\,\psi(x)(dx)^\l\right\}=
\left(\nu\,\phi(x)\psi(x)'-\l\,\phi(x)'\psi(x)\right)(dx)^{\nu+\l+1},
\label{Sch}
\end{equation}
where  $x$ is a local coordinate on $M$ and we identify tensor
densities with functions.
The operator (\ref{Sch}) is nothing but the Poisson bracket on $T^*S^1$ (or $T^*\bbR^1$). 

For every $(\nu,\l)\not=(0,0)$, the operator (\ref{Sch}) is the only $\Diff(S^1)$- (or
$\Diff(\bbR^1)$-) invariant operator, otherwise there are two linearly independent operators:
$\phi d(\psi)$ and $d(\phi)\psi$, where $d$ is the de Rham differential.

\medskip
\item 
There exist second order operators
$\cF_{\nu}\otimes\cF_{\l}\to\cF_{\nu+\l+2}$ given by the compositions:
\begin{equation}
\begin{array}{rcl}
\phi\otimes\psi&\mapsto&
\{d\phi,\,\psi\} \qquad \hbox{for} \quad \nu=0,\\[6pt]
\phi\otimes\psi&\mapsto&
\{\phi,\,d\psi\} \qquad \hbox{for}  \quad \l=0,\\[6pt]
\phi\otimes\psi&\mapsto&
d\,\{\phi,\,\psi\} \qquad \hbox{for}  \quad \nu+\l=-1.
\end{array}
\label{Sec1}
\end{equation}

\medskip

\item 
Three third-order bilinear invariant differential operators
$\cF_{\nu}\otimes\cF_{\l}\to\cF_{\nu+\l+3}$ are also given by compositions:
\begin{equation}
\begin{array}{rcl}
\phi\otimes\psi&\mapsto&
\{d\phi,\,d\psi\} \qquad \hbox{for} \quad (\nu,\l)=(0,0),\\[6pt]
\phi\otimes\psi&\mapsto&
d\,\{d\phi,\,\psi\} \qquad \hbox{for} \quad (\nu,\l)=(0,-2),\\[6pt]
\phi\otimes\psi&\mapsto&
d\,\{\phi,\,d\psi\} \qquad \hbox{for} \quad (\nu,\l)=(-2,0).
\end{array}
\label{compos3Op}
\end{equation}

\medskip

\item 
The only differential operator of order 3 which is not a composition of the
operators of lesser orders is the famous Grozman operator 
$G:\cF_{-\frac{2}{3}}(S^1)\otimes\cF_{-\frac{2}{3}}(S^1)\to\cF_{\frac{5}{3}}(S^1)$
already mentioned in Introduction, see (\ref{GroOpDef}). It is given by the following
expression:
\begin{equation}
G\left(
\phi(x)(dx)^{-\frac{2}{3}},\psi(x)(dx)^{-\frac{2}{3}}
\right)=
\left(2\left|
\matrix{
\phi(x)&\psi(x)\hfill\cr
\phi'''(x)&\psi'''(x)\hfill\cr
}
\right|+
3\left|
\matrix{
\phi'(x)&\psi'(x)\hfill\cr
\phi''(x)&\psi''(x)\hfill\cr
}
\right|
\right)(dx)^\frac{5}{3}.
\label{Gro}
\end{equation}
The $\Diff(S^1)$-invariance of this operator can be easily checked directly.

\begin{rmk}
{\rm
The operator (\ref{Gro}) remains one of the most mysterious invariant differential
operators. Its geometric and algebraic meaning was discussed in \cite{FF1,Fuc}.
}
\end{rmk}

\end{enumerate}

We will use the above bilinear operators to construct the symmetries, but we do not use
Grozman's classification result in our proof.

\subsubsection{Invariant projections from $\cD^k_{\l,\m}(S^1)$ to
$\cF_\nu$}\label{projectors}

Let us now give the list of $\Diff(S^1)$-invariant
linear maps from $\cD^k_{\l,\m}(S^1)$ to the space $\cF_\nu$.

\begin{enumerate}

\item

The well-known projection is the \textit{principal symbol} map
$
\s:\cD^k_{\l,\m}(S^1)\to\cF_{\m-\l-k}.
$
given by the expression
\begin{equation}
\label{sigma}
\s\left(
\sum_{i=0}^ka_i(x)\frac{d^i}{dx^i}
\right)=a_k(x)\,(dx)^{\m-\l-k}.
\end{equation}
The map $\s$ is obviously $\Diff(S^1)$-invariant for all $(\l,\m)$.

\medskip

\item
For all $(\l,\m)$, define a linear map
$$
V:\cD^k_{\l,\m}(S^1)\to\cF_{\m-\l-k+1}
$$
as follows:
\begin{equation}
\label{tp1}
V(A)=
 \left(
\a\,a'_k(x)
+ \b\,a_{k-1}(x)
\right)\,(dx)^{\m-\l-k+1}.
\end{equation}
where
$$
\a=\l\,k + \frac{k(k - 1)}{2},
\qquad
\b=\m-\l-k
$$
It is easy to check that this map is $\Diff(S^1)$-invariant.
The map (\ref{tp1}) is a ``first-order analog'' of the principal symbol.

\begin{rmk}
{\rm
If $\l+\m=1$, then this map is proportional to the principal symbol of the
$(k-1)$-th order operator $A-(-1)^k\,A^*$. In other words, 
$$
V=\s\circ(\Id-(-1)^kC)
$$
if $\l+\m=1$. 
}
\end{rmk}

\medskip

\item
In the particular case,
\begin{equation}
\label{WilMod}
\l=\frac{1-k}{2},
\qquad
\m=\frac{1+k}{2}
\end{equation}
The map (\ref{tp1}) vanishes. In this case, there are \textit{two} independent
projections onto $\cF_1$:
$$
A\mapsto{}a'_k(x)\,dx,
\qquad
A\mapsto{}a_{k-1}(x)\,dx
$$
which are $\Diff(S^1)$-invariant.

\medskip

\item
It turns out that for some special values of the parameters $\l$ and $\m$, 
there exist second-order analogues of the operators (\ref{sigma}) and (\ref{tp1}).

\begin{prop}
For every $k\geq 3$ and $(\l, \m)$ satisfying the relation
\begin{equation}
\label{hk}
\left(\l+\frac{k-2}{3}\right)
\left(\m-\frac{k+1}{3}\right)+
\frac{1}{36}(k+1)(k-2)\,=\,0,
\end{equation}
there exists a $\Diff(S^1)$-invariant map 
$W: \cD^k_{\l,\,\m}\to\cF_{\m-\l-k+2}$ given by
\begin{equation}
\label{maphk}
W(A)=
 \left(
 \a_2\,a_k''(x)
+\a_1\,a_{k-1}'(x)
+\a_0\,a_{k-2}(x)
\right)\,(dx)^{\m-\l-k+2},
\end{equation} 
where the coefficients are defined by
\begin{equation}
\label{W}
\begin{array}{lll}
\a_2 & = & \frac{2}{3} k (k - 1) (k + 3 \lambda - 2)^2\\[4pt]
\a_1 & = & 2 (k - 1 )(k + 3 \lambda - 2)(2 - 2\lambda - k)\\[4pt]
\a_0 & = & 3 k^2 + 12 \lambda k + 12 \lambda^2 - 11 k - 24 
\lambda +10.
\end{array}
\end{equation} 
\end{prop}

\begin{proof}
Straightforward.
\end{proof}

\medskip

\item
There exists one more invariant differential projection
$\cD^k_{0,1}(S^1)\to\cF_0(S^1)$ given by the composition
\begin{equation}\label{specproj}
\pi_{\delta}=P_0\circ{}C\circ\d^{-1}\circ(\Id- P_{0}),
\end{equation}
where $P_{0}$ is defined by (\ref{pro}) and $C$ is the conjugation.

\medskip

\item
If there is an invariant map from $\cD^k_{\l,\m}(S^1)$ to the space of 1-forms
$\cF_{1}$, then one can integrate the result and
obtain  a \textit{non-local} (i.e., non-differential) invariant linear map with values in
$\bbR\subset\cF_0$.  For instance, for $\cD^k_{0,1}(S^1)$, the
operator $P_0$ defined by~(\ref{pro}) satisfies the required condition. One gets a
$\Diff(S^1)$-invariant map:
\begin{equation}
\label{nonloc}
A\mapsto\int_{S^1}a_0(x)\,dx,
\qquad
A\in\cD^k_{0,1}(S^1).
\end{equation}

\end{enumerate}

It was proven in \cite{MAT1} that there are no other $\Diff(S^1)$-invariant projections
$\cD^k_{\l,\m}(S^1)\to\cF_\nu$ than the above ones and their compositions
with $C$ and $d$. We use these operators to construct the generators of the symmetry
algebras (but we do not use the classification result of \cite{MAT1} in the proofs of our
theorems).

\section{Computing the algebras of symmetry}\label{ComputingSec}

We will now investigate, case by case, the non-trivial algebras $\cI^k_{\l,\m}(S^1)$. We
will construct the generators of these algebras and calculate the multiplication tables. We
then give an explicit identification of the algebras
$\cI^k_{\l,\m}(S^1)$ with the matrix algebras introduced in Section \ref{martix}.
The constructions of this section prove that the algebras $\cI^k_{\l,\m}(S^1)$ are 
\textit{at least} as big as stated in Section \ref{MainResSec}.

The proof of the second part of our classification theorems, namely that there are 
\textit{no other} symmetries than we construct and study here, will be given in Section
\ref{EndProof}. 

\subsection{The algebra $\cI^k_{0,1}(S^1)$}\label{explicitBis}

Let us start with the most particular algebra $\cI^k_{0,1}(S^1)$ for all $k$. 
One has in this case
$$
\cI^k_{0,1}(S^1)=
\mathrm{Span}
\left(
\Id,C,P_0,P_0^*,P_1,L
\right)
$$
where the generators are defined by (\ref{CA})-(\ref{leP1}).

Let us now calculate the
relations between these generators.

\begin{prop}
\label{prop1}
The multiplication table for the associative algebra $\cI^k_{0,1}(S^1)$ is as follows:
\begin{equation}
\label{table}
\begin{array}{c||c|c|c|c|c|c|}
 & \Id & P_0& C & P_0^* & P_1 &L\\[4pt]\hline\hline
\Id & \Id & P_0& C & P_0^* & P_1 &L\\[4pt]\hline
P_0 &P_0&P_0&P_0^*&P_0^*&0&0\\[4pt]\hline
C&C&P_0&\Id&P_0^*&P_0^*\!-\!P_1\!-\!P_0&-L\\[4pt]\hline
P_0^*&P_0^*&P_0&P_0&P_0^*&P_0^*-P_0&0\\[4pt]\hline
P_1&P_1&0&-P_1&0&P_1&L\\[4pt]\hline
L&L&L&L&L&0&0\\\hline
\end{array}
\end{equation}
\end{prop}
\begin{proof}
First, consider the product of $P_1$ and $C$. From the definition
(\ref{CA}) one obtains
$$
\begin{array}{rcl}
(P_1\,{}C)(A) &=& 
\displaystyle
P_1\left(
\sum_{i=0}^k(-1)^i\left(\frac{d}{d{}x}\right)^i\circ{}a_i(x)
\right)
\\[8pt]
&=&
\displaystyle
P_1\left(
\sum_{i=0}^k\sum_{j=0}^i
(-1)^i\,{i\choose{}j}\,a_i^{(i-j)}(x)\,
\frac{d^j}{d{}x^j}
\right)
\\[8pt]
&=&
\displaystyle
P_1\left(
\sum_{j=0}^k\sum_{i=j}^k
(-1)^i\,{i\choose{}j}\,a_i^{(i-j)}(x)\,
\frac{d^j}{d{}x^j}
\right)
\end{array}
$$
and then from (\ref{leP1}) it follows that
$$
\qquad\qquad\qquad
\begin{array}{rcl}
&=&
\displaystyle
\left(
\sum_{j=1}^k\sum_{i=j}^k
(-1)^{i+j-1}\,{i\choose{}j}\,a_i^{(i-1)}(x)\,
\right)\circ{}d\\[12pt]
&=&
\displaystyle
\left(
\sum_{i=1}^k\sum_{j=1}^i
(-1)^{i+j-1}\,{i\choose{}j}\,a_i^{(i-1)}(x)\,
\right)\circ{}d\\[12pt]
&=&
\displaystyle
\left(
\sum_{i=1}^k
(-1)^{i}\,a_i^{(i-1)}(x)\,
\right)\circ{}d\\[12pt]
&=&-P_1(A)
\end{array}
$$
as presented in table (\ref{table}).

Now, consider the product $C\,P_1$. One then has from (\ref{CA}),
(\ref{leP1}) and (\ref{leP0'expl}):
$$
\begin{array}{rcl}
(C\,P_1)(A)
&=&
\displaystyle
\sum_{i=1}^k(-1)^i\,\left(
a_i^{(i-1)}(x)\,
\frac{d}{d{}x}
+\,a_i^{(i)}(x)\right)\\[14pt]
&=&
P_0^*-P_1-P_0.
\end{array}
$$
Furthermore, one obtains
$P_0^*\,P_1=(P_0\,C\,P_1)(A)=P_0^*-P_0$.

For other products of the the generators the results given in the table immediately follow
from the definition.
\end{proof}

Let us finally give an explicit isomorphism between the algebra $\cI^k_{0,1}(S^1)$ and
the matrix algebra $\fb\oplus\bbR^2$ described in Section \ref{martix}.

One checks using the multiplication table (\ref{table}) that the formul{\ae}
$$
\begin{array}{rcl}
\bar{a}&=&\half(2P_1+P_0-P_0^*),\\[6pt]
\bar{b}&=&\half(P_0+P_0^*),\\[6pt]
\bar{c}&=&\half(P_0-P_0^*),\\[6pt]
\bar{d}&=&L,
\end{array}
$$
where $\bar{a},\bar{b},\bar{c},\bar{d}$ are the generators of $\fb$, see Appendix
\ref{Appen}, define an isomorphism of the associative algebras
$$
\mathrm{Span}(P_0,P_0^*,P_1,L)
\cong\fb.
$$
The two more generators
$$
\begin{array}{rcl}
z_1&=&\Id+C-P_0-P_0^*\\[6pt]
z_2&=&\Id-C-P_0+P_0^*-2P_1
\end{array}
$$
are in the center and span the second summand $\bbR^2$.

The above generators are linearly independent if $k\geq3$ and span the algebra
$$
\cI^k_{0,1}(S^1)\cong\fb\oplus\bbR^2,
$$
in accordance with Theorem
\ref{Dim1Thm1}, 3, Theorem
\ref{Dim1Thm2}, 4 and Theorem \ref{Dim1Thm3}, 7.

If $k=2$, then $z_2=0$ so that $\cI^2_{0,1}(S^1)\cong\fb\oplus\bbR$, see Theorem
\ref{Dim1Thm4}, 5. Finally, if $k=1$, then $z_1=z_2=0$ and one has
$\cI^1_{0,1}(S^1)\cong\fb$ as stated Theorem \ref{Dim1Thm5}, 2.

\subsection{Algebras of symmetry in order $k\geq5$}

Assume that $k\geq5$. We already investigated the algebra $\cI^k_{0,1}(S^1)$.
There are two more non-trivial algebras in this case, namely the algebra $\cI^k_{0,0}(S^1)$
and the algebra
$\cI^k_{1,1}(S^1)$ which is isomorphic to $\cI^k_{0,0}(S^1)$ by conjugation.

The algebra of symmetry $\cI^k_{0,0}(S^1)$ is as follows
$$
\cI^k_{0,0}(S^1)=\mathrm{Span}
\left(
\Id,P_0,S
\right),
$$
where $P_0$ and $S$ are as in (\ref{pro}) and (\ref{Tdelta}), respectively. These
operators are independent for $k\geq4$. We have:
$$
P_0\,S=S\,P_0=P_0,
\qquad
P_0^2=P_0,
\qquad
S^2=\Id.
$$
Indeed, the first two relations are due to the fact that the scalar term of $S(A)$ is equal
to $a_0(x)$, cf. eq. (\ref{Tdelta}) and the explicit expression for $S$.
Put 
$$
1=\Id,\quad\bar{a}_1=P_0
\quad \hbox{and}\quad
\bar{a}_2=\frac{1}{\sqrt{2}}(\Id-S).
$$
One obtains the
generators of the algebra $\bbR^3$, cf. Appendix \ref{Appen}. Finally, one has
$$
\cI^k_{0,0}(S^1)\cong\bbR^3,
$$
as stated in
Theorem \ref{Dim1Thm1}, 2.

The algebras $\cI^k_{0,\m}(S^1)$ corresponding to the generic values of $\m$ have
only two generators: $\Id$ and $P_0$. These algebras are obviously isomorphic to $\bbR^2$.

\subsection{Algebras of symmetry in order 4}

Consider the modules of differential operators of order $k=4$:
$$
A=a_4(x)\frac{d^4}{dx^4}
+a_3(x)\frac{d^3}{dx^3}
+a_2(x)\frac{d^2}{dx^2}+
a_1(x)\frac{d}{dx}+
a_0(x).
$$
We will study all the exceptional modules systematically and investigate every
non-trivial algebra of symmetry.

The generators of the algebras $\cI^4_{0,1}(S^1)$ and $\cI^4_{0,0}(S^1)\cong\cI^4_{1,1}(S^1)$
are the same as for $k=5$. Let us consider other interesting cases.

\subsubsection{The algebras $\cI^4_{0,\frac{5}{4}}(S^1)$ 
and $\cI^4_{-\frac{1}{4},1}(S^1)$}

We already constructed two generators of the algebra $\cI^4_{0,\frac{5}{4}}(S^1)$, 
namely $\Id$ and $P_0$. One extra generator is obtained by the following procedure.

The values $(\l,\m)=(0,\frac{5}{4})$ satisfy the relation (\ref{hk}), so
that the map 
$$
W:\cD^4_{0,\frac{5}{4}}(S^1)\to\cF_{-\frac{3}{4}}
$$
defined by (\ref{maphk}) is $\Diff(S^1)$-invariant. There is a second-order bilinear
differential operator
$$
J:\cF_{-\frac{3}{4}}\otimes\cF_{0}\to\cF_{\frac{5}{4}}
$$
defined by the second formula in (\ref{Sec1}). Applying the construction of Section
\ref{SBOTDSec} we consider the composition $J\circ{}W$ defined as in Section \ref{SBOTDSec}
 to obtain an element of algebra
$\cI^4_{0,\frac{5}{4}}(S^1)$.

Let us now compute the relations between the generators.
The constructed map is given by the following explicit formula
\begin{equation}
\label{ExpSymNO}
\begin{array}{rcl}
\left(J\circ{}W\right)(A)&=&
\displaystyle\left(
\frac{16}{7}a_4''(x)-\frac{12}{7}a_3'(x)+a_2(x)
\right)\frac{d^2}{dx^2}\\[12pt]
&&+
\displaystyle\frac{4}{3}
\left(
\frac{12}{7}a_4'''(x)-\frac{9}{7}a_3''(x)+\frac{3}{4}a_2'(x)
\right)\frac{d}{dx}.
\end{array}
\end{equation}
Note that the right hand side is understood as an element of $\cD^4_{0,\frac{5}{4}}(S^1)$.
Since $P_0(A)=a_0(x)$, the product of $J\circ{}W$ and $P_0$ vanishes:
$$
\left(J\circ{}W\right)\,P_0=P_0\,\left(J\circ{}W\right)=0.
$$
One also has the relations 
$$
P_0^2=P_0
\qquad\hbox{and}\qquad
\left(J\circ{}W\right)^2=J\circ{}W.
$$

Finally, one gets the following answer:
$$
\cI^4_{0,\frac{5}{4}}(S^1)=
\mathrm{Span}
\left(
\Id,P_0,J\circ{}W
\right)\cong\bbR^3,
$$
as stated by Theorem \ref{Dim1Thm2}, 2. 

The conjugation establishes an isomorphism between the algebras $\cI^4_{-\frac{1}{4},1}(S^1)$
and $\cI^4_{0,\frac{5}{4}}(S^1)$.

\subsubsection{The algebras $\cI^4_{0,3}(S^1)$ and $\cI^4_{-2,1}(S^1)$}

The algebra $\cI^4_{0,3}(S^1)$ has the generators $\Id,P_0$ and the following one
constructed in Section \ref{SBOTDSec}. 

Consider the projection
$V:\cD^4_{0,3}(S^1)\to\cF_0$ defined by formula (\ref{tp1}) and the
third-order bilinear map $J:\cF_0\otimes\cF_0\to\cF_3$, namely the first of the three
operators (\ref{compos3Op}). Their composition 
$
J\circ{}V
$
is an element of $\cI^4_{0,3}(S^1)$:
\begin{equation}
\label{ExpSymNOdn}
\left(J\circ{}V\right)(A)=
\left(6a_4''(x)-a_3'(x)
\right)
\frac{d^2}{dx^2}
-
\left(
6a_4'''(x)-a_3''(x)
\right)\frac{d}{dx},
\end{equation}
where the right hand side is understood as an element of $\cD^4_{0,3}(S^1)$.

Finally, the algebra $\cI^4_{0,3}(S^1)$ is of the form
$$
\cI^4_{0,3}(S^1)=
\mathrm{Span}
\left(
\Id,P_0,J\circ{}V
\right).
$$
To obtain the isomorphism $\cI^4_{0,3}(S^1)\cong\fa\oplus\bbR$ (see Theorem \ref{Dim1Thm2}
part 3), one checks the following relations
$$
P_0\,\left(J\circ{}V\right)=\left(J\circ{}V\right)\,P_0=(J\circ{}V)^2
=0.
$$
Then the standard generators of $\fa\oplus\bbR$ correspond to $\{\Id-P_0,J\circ{}V,P_0\}$.

The algebra $\cI^4_{-2,1}(S^1)$ is isomorphic to $\cI^4_{0,3}(S^1)$ by conjugation.

\subsubsection{The algebra
$\cI_{-\frac{2}{3},\frac{5}{3}}^4(S^1)$
and the Grozman operator}\label{explicitGro}

The conjugation map $C$ and $\Id$ are, of course, generators of symmetry of the
module $\cD_{-\frac{2}{3},\frac{5}{3}}^4(S^1)$. One extra generator can be obtained as
follows.

Consider the operator (\ref{tp1})
$$
V:\cD_{-\frac{2}{3},\frac{5}{3}}^4(S^1)\to\cF_{-\frac{2}{3}}
$$ 
and compose it with the Grozman operator $G$ given by (\ref{Gro}); we 
obtain (up to a constant) the following operator:
\begin{equation}
\label{SecGen}
\begin{array}{rcl}
\left(G\circ V\right) (A)&=&
\displaystyle
\left(a_{3}(x)-2\,a_{4}'(x) \right)\frac{d^3}{d x^3} 
+ \left(\frac{3}{2}\,a_{3}'(x)-3\,a_{4}''(x) \right)\frac{d^2}{d x^2} \\[12pt]
&&
\displaystyle
\hskip1cm-
\left(\frac{3}{2}\,a_{3}''(x)-3\,a_{4}'''(x) \right)\frac{d}{d x} -
\left(a_{3}'''(x)-2\,a_{4}^{(IV)}(x) \right)
\end{array}
\end{equation}
which is a generator of $\cI_{-\frac{2}{3},\frac{5}{3}}^4(S^1)$.

The relations between the conjugation map and the above operator are:
$$
\left(G\circ V\right)\,C=C\,\left(G\circ V\right)=-G\circ V.
$$
One also has:
$$
\left(G\circ V\right)^2=G\circ V.
$$

One easily deduces from the above relations that the algebra
$$
\cI_{-\frac{2}{3},\frac{5}{3}}^4(S^1)=
\mathrm{Span}
\left(
\Id,C,G\circ{}V
\right)
$$
is, indeed, isomorphic to $\bbR^3$, cf. Theorem \ref{Dim1Thm2}, 2.

\subsection{Algebras of symmetry in order 3}

Consider the differential operators of order $k=3$:
$$
A=a_3(x)\frac{d^3}{dx^3}
+a_2(x)\frac{d^2}{dx^2}+
a_1(x)\frac{d}{dx}+
a_0(x).
$$
We will describe all the non-trivial algebras of symmetry.

\subsubsection{The algebra
$\cI_{-\frac{2}{3},\frac{5}{3}}^3(S^1)$}\label{explicit}

The conjugation map $C$, as well as the
identity $\Id$, are, of course, generators of the symmetry algebra
$\cI_{-\frac{2}{3},\frac{5}{3}}^3(S^1)$. Let us construct one more generator. The principal
symbol map is of the form:
$$
\s:\cD_{-\frac{2}{3},\frac{5}{3}}^3(S^1)\to\cF_{-\frac{2}{3}}.
$$
We compose it with the Grozman operator to obtain a new generator $G\circ\s$.
This operator is given by the same formula (\ref{SecGen}) as above, but with
$a_4(x)\equiv0$.
The relations between the generators are also the same as above, so that the symmetry algebra
is
$
\cI_{-\frac{2}{3},\frac{5}{3}}^3(S^1)=\bbR^3.
$

\subsubsection{The hyperbola $(3\l+1)(3\m-4)=-1$}\label{HypSec}

Consider the class of modules $\cD^3_{\l,\m}(S^1)$ with $(\l,\m)$ satisfying
the quadratic relation
\begin{equation}
\label{WkThree}
(3\l+1)(3\m-4)=-1,
\end{equation} 
see Theorem \ref{Dim1Thm3}, 1.

First of all, we observe that this relation is precisely the relation (\ref{hk}) specified
for $k=3$. The operator
$
W:\cD^3_{\l,\m}(S^1)\to\cF_{\m-\l-1}
$
 is then well-defined.
Composing this operator with the Poisson bracket (\ref{Sch})
$$
\{\cdot,\cdot\}:\cF_{\m-\l-1}\otimes\cF_{\l}\to\cF_{\m},
$$
one obtains a generator of the algebra $\cI^3_{\l,\m}(S^1)$. Let us denote this generator by
$\mathcal W$:
\begin{equation}
\label{FirGen}
\begin{array}{rcl}
{\mathcal W}(A)&=&
\displaystyle
(\m-\l-1)
\left(\a_2\,a_{3}''(x) + \a_1\,a_{2}'(x) + 
\a_0\,a_{1}(x)\right) 
\frac{d}{d x} 
\\[12pt]
&&
\displaystyle
- \l
\left(\a_2\,a_{3}'''(x) + \a_1\,a_{2}''(x) + 
\a_0\,a_{1}'(x)
\right),
\end{array}
\end{equation}
where according to (\ref{W})
$$
\begin{array}{lll}
\a_2 & = & (3\l+1)^2,\\[4pt]
\a_1 & = & -(3\l+1)(1-2\l),\\[4pt]
\a_0 & = & 3\l^2+3\l+1.
\end{array}
$$

In the generic case, $(\l,\m)\not=(0,1)$ or $(-\frac{2}{3},\frac{5}{3})$ the symmetry algebra
has two generators:
$
\cI^3_{\l,\m}(S^1)=
\mathrm{Span}
\left(
\Id,S
\right).
$
The generator $\mathcal W$ satisfies the relation 
$$
{\mathcal W}^2=\a_0(\m-\l-1){}{\mathcal W}.
$$
But $\a_0\neq0$ and if $\m-\l-1=0$, then (\ref{WkThree}) implies $(\l,\m)=(0,1)$. Finally,
in the generic case, one obtains:
$$
\cI^3_{\l,\m}(S^1)\cong\bbR^2.
$$

\begin{rmk}
{\rm
The module $\cD^3_{-\frac{2}{3},\frac{5}{3}}(S^1)$ belongs to the family (\ref{WkThree}). 
We have already considered this module separately, see Section \ref{explicit}.
In this case, we have three generators:
$
\cI^3_{-\frac{2}{3},\frac{5}{3}}(S^1)=
\mathrm{Span}
\left(
\Id,C,\mathcal W
\right)
$
which are different from $G\circ\s$. One checks, however, that
the generator $G\circ\s$ can be expressed in terms of the above ones:
$$
G\circ\s=\half(\Id-C)-\frac{9}{4}{\mathcal W}.
$$ 
}
\end{rmk}

\subsubsection{The line $\m-\l=2$}\label{ExLineSec}

Consider the family of modules $\cD^3_{\l,\m}(S^1)$ satisfying the property $\m-\l=2$ as in
Theorem \ref{Dim1Thm3}, 2.

The operator (\ref{tp1}) is, in this case, $V:\cD^3_{\l,\m}(S^1)\to\cF_0$. Consider its
composition with the second-order bilinear operator $J:\cF_0\otimes\cF_\l\to\cF_\m$ given by
the first formula in (\ref{Sec1}). In the generic case, that is, where 
$$
\textstyle
\left(\l,\m\right)\neq\left(0,2\right),
\,\left(-\frac{1}{2},\frac{3}{2}\right),
\,\left(-1,1\right),
$$
one has
$
\cI^3_{\l,\m}(S^1)=
\mathrm{Span}
\left(
\Id,J\circ{}V
\right).
$

One obtains the following explicit formula for the
constructed generator:
\begin{equation}
\label{LineAddGen}
\left(J\circ{}V\right)(A)=
\left(3(\l+1)\,a_{3}''(x) -a_{2}'(x)\right)\frac{d}{d x} 
-
\l\left(3(\l+1)\,a_{3}'''(x) -a_{2}''(x)\right)
\end{equation}
and immediately gets the following relation:
$$
\left(J\circ{}V\right)^2=0.
$$
This implies $\cI^3_{\l,\m}(S^1)\cong\fa$ in the generic case.

\subsubsection{The modules
$\cD^3_{0,2}(S^1),\,\cD^3_{-1,1}(S^1)$
and $\cD^3_{-\frac{1}{2},\frac{3}{2}}(S^1)$}\label{ExeptLineSec}

Let us consider some exceptional modules still satisfying $\m-\l=2$.

In the case, $\left(\l,\m\right)=\left(0,2\right)$, one has an extra generator of symmetry,
as compared with the preceding section. It is given by the operator $P_0$ as in
(\ref{pro}). One then has from (\ref{LineAddGen})
$$
\left(J\circ{}V\right)P_0=0,
\quad
P_0\left(J\circ{}V\right)=0.
$$
The isomorphism
$$
\cI^3_{0,2}(S^1)=\mathrm{Span}
\left(
\Id,J\circ{}V,P_0
\right)\cong\fa\oplus\bbR
$$ 
is then obvious in accordance with Theorem \ref{Dim1Thm3}, 4. 

The conjugation map $C$ establishes an isomorphism
$\cI^3_{-1,1}(S^1)\cong\cI^3_{0,2}(S^1)$, so that tha algebra $\cI^3_{-1,1}(S^1)$ is also
isomorphic to $\fa\oplus\bbR$.

In the interesting case $\cD^3_{-\frac{1}{2},\frac{3}{2}}(S^1)$, the extra generator is
given by the conjugation map $C$, so that
$\cI^3_{-\frac{1}{2},\frac{3}{2}}(S^1)=\mathrm{Span}
\left(
\Id,J\circ{}V,C
\right).$
The relations between the generators are
$$
\left(J\circ{}V\right)\,C=J\circ{}V,
\qquad
C\,\left(J\circ{}V\right)=-J\circ{}V
$$
as follows from (\ref{LineAddGen}) and (\ref{CA}).
One obtains 
$$
\cI^3_{-\frac{1}{2},\frac{3}{2}}(S^1)\cong\ft_2,
$$
as stated by Theorem
\ref{Dim1Thm3}, 5.

\subsubsection{The modules
$\cD^3_{0,3}(S^1)$
and $\cD^3_{-2,1}(S^1)$}\label{LastExTri}

The principal symbol map (\ref{sigma}) is as follows $\s:\cD^3_{0,3}(S^1)\to\cF_0$. Compose
this map with the third-order bilinear operator $J:\cF_0\otimes\cF_0\to\cF_3$ defined by the
first equation in (\ref{compos3Op}). The explicit expression of the constructed generator is
\begin{equation}
\label{TheLastGen}
\left(J\circ\s\right)(A)=
a_{3}'(x)\frac{d^2}{dx^2}
-
a_{3}''(x)\frac{d}{dx}.
\end{equation}
The symmetry algebra is then
$\cI^3_{0,3}(S^1)=\mathrm{Span}
\left(
\Id,J\circ\s,P_0
\right).$
One easily gets the relations 
$$
\left(J\circ\s\right)P_0=P_0\left(J\circ\s\right)=\left(J\circ\s\right)^2=0
$$ 
and, finally, $\cI^3_{0,3}(S^1)\cong\fa\oplus\bbR$, see Theorem
\ref{Dim1Thm3}, 4. The algebra $\cI^3_{-2,1}(S^1)$ is isomorphic to the above one by
conjugation.

\subsection{Algebras of symmetry in orders 2}\label{SecOrdSymSec}

Consider now the modules of differential operators of order 2. 

For all $(\l,\m)$, there is
a generator of the algebra $\cI^2_{\l,\m}(S^1)$ given by the composition of the projection
(\ref{tp1})
$$
V:\cD^2_{\l,\m}(S^1)\to\cF_{\m-\l-1}
$$
and the Poisson bracket
$$
\{\cdot,\cdot\}:\cF_{\m-\l-1}\otimes\cF_\l\to\cF_\m.
$$
The explicit formula for this generator is as follows:
\begin{equation}
\label{FirNNewGen}
\begin{array}{rcl}
{\mathcal V}(A)&=&
\displaystyle
(\m-\l-1)
\left((2\l+1)\,a_{2}'(x) + 
(\m-\l-2)\,a_{1}(x)\right) 
\frac{d}{dx} 
\\[12pt]
&&
\displaystyle
- \l
\left((2\l+1)\,a_{2}''(x) + 
(\m-\l-2)\,a_{1}'(x)\right).
\end{array}
\end{equation}

This generator satisfies the following relation:
$$
{\mathcal V}^2=(\m-\l-1)(\m-\l-2)\,{\mathcal V}.
$$
In the generic case, the algebra of symmetry is 
$\cI^2_{\l,\m}(S^1)=\mathrm{Span}
\left(
\Id,{\mathcal V}
\right)$.
It is obviously isomorphic to $\bbR^2$.

If either $\m-\l=1$ or $\m-\l=2$, then
${\mathcal V}^2=0$. In these cases, $\cI^2_{\l,\m}(S^1)=\mathrm{Span}
\left(
\Id,{\mathcal V}
\right)\cong\fa$, see Theorem \ref{Dim1Thm4}, 1.

Of course, if $\l=0$, then there is one more generator, namely $P_0$. Then one has
$$
P_0\,{\mathcal V}={\mathcal V}\,P_0=0
$$
and so $\cI^2_{0,\m}(S^1)=\mathrm{Span}
\left(
\Id,{\mathcal V},P_0
\right)$ which is isomorphic to $\bbR^3$ for generic $\m$ while
$$
\cI^2_{0,2}(S^1)\cong\cI^2_{-1,1}(S^1)\cong\fa\oplus\bbR
$$
see Theorem \ref{Dim1Thm4}, 2 and \ref{Dim1Thm4}, 4.

In the case where $\l+\m=1$, the conjugation map is well-defined. One checks that in this
case the generator (\ref{FirNNewGen}) is a linear combination of $\Id$ and $C$:
$$
{\mathcal V}=\l(2\l+1)\left(C-\Id\right)
$$
so that $\cI^2_{\l,\m}(S^1)\cong\bbR^2$.

The exceptional case $(\l,\m)=(-\frac{1}{2},\frac{3}{2})$ corresponds to (\ref{WilMod}).
There are two invariant projections from $\cD^2_{-\frac{1}{2},\frac{3}{2}}(S^1)$ to $\cF_1$
in this case. Composing one of them with the Poisson bracket
$$
\{\cdot,\cdot\}:\cF_1\otimes\cF_{-\frac{1}{2}}\to\cF_{\frac{3}{2}}
$$
one obtains a generator
$$
a_2(x)\frac{d^2}{dx^2}+a_1(x)\frac{d}{dx}+a_0(x)
\mapsto{}a_2'(x)\frac{d}{dx}+\frac{1}{2}\,a_2''(x)
$$
independent of $\Id$ and $C$. One easily gets an isomorphism
$\cI^2_{-\frac{1}{2},\frac{3}{2}}(S^1)\cong\ft_2$.

\section{Proof of the main theorems}\label{EndProof}

In this section we prove that there are no other
symmetries of the modules $\cD^k_{\l,\m}(S^1)$ than those constructed above. In other
words, we give here a complete classification of symmetries.

Let $T$ be a linear map (\ref{Endom}) commuting with the $\Diff(S^1)$-action. 
There are two cases:

\begin{enumerate}
\item
The map $T$ is \textit{local}, that is, one has
$\Supp(T(A))\subset\Supp(A)$  for all
$A\in\cD^k_{\l,\m}(S^1)$. In this case, the famous Peetre theorem (see \cite{Pee})
guarantees that $T$ is a \textit{differential operator} in coefficients of $A$.

\medskip
\item

The map $T$ is \textit{non-local}, that is, for some $A\in\cD^k_{\l,\m}(S^1)$
vanishing in an open subset $U\subset{}S^1$, the operator $T(A)$ does not vanish on $U$.
\end{enumerate}
These two cases are completely different and should be treated separately.

\subsection{The identification}\label{IdiotSec}

Let us fix a parameter $x$ on $S^1$ and the corresponding coordinate $\xi$ on the fibers of
$T^*{S^1}$.

For our computations, we will need to identify the spaces $\cD_{\l,\m}(S^1)$ and
$\cS_{\d}(S^1)$ using the map
\begin{equation}
\label{Identif}
\s_{\mathrm{tot}}:
\cD_{\l,\m}(S^1)\to
\cS_{\d}(S^1)
\end{equation}
assigns to an operator (\ref{DiOpA}) the polynomial
on $T^*S^1$ given by (\ref{SymDiOpA}). 

The map $\s_{\mathrm{tot}}$ is an isomorphism of vector spaces but {\it not} an
isomorphism of $\Diff(S^1)$-modules. It will, nevertheless, allow us to
compare the $\Diff(S^1)$-action on both spaces.

\subsection{The affine Lie algebra}\label{StatSec}

We introduce our main tool that will allow un to use the results of the classic 
invariant theory.

Let $x$ be an affine parameter on $S^1$, the
Lie algebra $\aff$ of affine transformations is the two-dimensional Lie algebra generated by
the translations and linear vector fields:
\begin{equation}
\aff=
\mathrm{Span}
\left(
\frac{d}{dx},\;
x\,\frac{d}{dx}
\right).
\label{aff}
\end{equation}
Since $\aff$ is a subalgebra of $\Vect(S^1)$,
every $\Diff(S^1)$-invariant map has to commute with the $\aff$-action.

\begin{prop}
\label{affPro}
The $\aff$-action on $\cD^k_{\l,\m}(S^1)$ depends only
on the difference $\m-\l$ and coincides with the action on $\cS_{\d}(S^1)$ after
identification $(\ref{Identif})$.
\end{prop}
\begin{proof}
Straightforward.
\end{proof}

The well-known result of invariant theory states that the associative algebra of
differential operators on $T^*S^1$ commuting with the $\aff$-action is generated by
\begin{equation}
E=
\xi\,\frac{\partial}{\partial{}\xi}
\quad
\hbox{and}
\quad
D=
\frac{\partial}{\partial{}x}\,\frac{\partial}{\partial{}\xi},
\label{ED}
\end{equation}
see  \cite{Wey}.
The operator $E$ is called the \textit{Euler field} and the operator $D$ the
\textit{divergence}.

Every differential $\Diff(M)$-invariant operator (\ref{Endom}) 
can therefore be expressed in terms of these operators. 
In local coordinates, any $\Diff(M)$-invariant map (\ref{Endom}) is therefore of the
form 
$$
T=T(E,D).
$$

\begin{exe}
{\rm
The expression
$$
C=\exp(D)\circ\exp(i\pi{}E)
$$
 for the conjugation map (\ref{CA}) is worth mentioning for the aesthetic reasons.
}
\end{exe}

\subsection{The local case: invariant differential operators}

We consider the algebra ${\cI^k_{\l,\m}}^{\rm loc}(S^1)$ of local (and thus
differential) $\Diff(S^1)$-invariant linear maps~(\ref{Endom}). 

Let us restrict the map $T$ to the homogeneous component $\cF_{k-\d}$ in (\ref{SymDiOpA}).
Since the Euler operator $E$ reduces to a constant, one has
\begin{equation}
\label{restr}
\left.
T
\right|_{\cF_{\d-k}}=\sum_{\ell=0}^{k}
T_{k,\ell}\,D^\ell
\end{equation}
where $T_{k,\ell}$ are some constants. 

The operator $T$ has to commute with the $\Vect(S^1)$-action on $\cD_{\l,\m}^k(S^1)$.
Consider the Lie subalgebra $\Sl(2)\subset\Vect(S^1)$ generated by three vector fields:
\begin{equation}
\label{sln+1}
\Sl(2)=
\mathrm{Span}
\left(
\frac{d}{d{}x},
\;
x\,\frac{d}{d{}x},
\;
x^2\,\frac{d}{d{}x},
\right).
\end{equation}

\begin{rmk}
\label{AffRmK}
{\rm
Assume that $x$ is an \textit{affine parameter} on $S^1$, that is, we identify $S^1$ with
$\bbRP^1$ with homogeneous coordinates $(x_1:x_2)$ and choose $x=x_1/x_2$. The vector fields
(\ref{sln+1}) are then globally defined and correspond to the standard projective structure
on $\bbRP^1$, see, e.g., \cite{KIR}. 
}
\end{rmk}

\begin{prop}
\label{slprop}
A linear map (\ref{Endom}) written in the form (\ref{restr}) is $\Sl(2,\bbR)$-invariant
if and only if it satisfies the recurrence relation
\begin{equation}
\label{relat}
\left(k+2\l-1\right)\,T_{k-1,\ell-1}-
\left(k+2\l-\ell\right)\,T_{k,\ell-1}-
\ell\left(2(\m-\l)-2k+\ell-1\right)\,T_{k,\ell}
=0.
\end{equation}
\end{prop}
\begin{proof}
The form (\ref{restr}) is already invariant with respect to the affine subalgebra
(\ref{aff}) of $\Sl(2,\bbR)$. It remains to impose the equivariance
condition with respect to the vector field $X=x^2\frac{d}{d{}x}$. 
Let us compute the Lie derivative (\ref{comm}) along this vector field. Again, we use the
identification (\ref{Identif}) and express it in terms of symbols. One has
$$
\cL^{\l,\m}_X=L_X^{\m-\l}-(2\l+E)\frac{\partial}{\partial\xi},
$$
where 
$$
L_X^{\m-\l}=
x^2\frac{\partial}{\partial{}x}
-2x\xi\frac{\partial}{\partial{}\xi}
+2(\m-\l)\,x.
$$
The equivariance condition $[T,\cL^{\l,\m}_X]=0$ readily leads to the relation
(\ref{relat}).
\end{proof}

Let us now impose the equivariance condition with respect to the vector field
$$
X=x^3\,\frac{d}{d{}x}.
$$
This vector field is not globally defined on $S^1$ and has a singularity at $x=\infty$.
Indeed, choose the coordinate $z=\frac{1}{x}$ in a vicinity of the point $z=0$,
the vector field $X$ is written $X=-\frac{1}{z}\frac{d}{dz}$, cf. Remark \ref{AffRmK}.
However, every $\Diff(S^1)$-invariant differential operator $T$ has to commute with this
vector field everywhere for $x\neq\infty$.

\begin{prop}
\label{slExtprop}
A differential operator $T$ commutes with the action of $X=x^3\,\frac{d}{d{}x}$
if and only if $T$ satisfies the relation (\ref{relat}) together with the relation
\begin{equation}
\label{LastEq}
\left(6\l+3k-3\right)\,T_{k-1,\ell-1}
+\ell\left(3(\m-\l)-3k+\ell-2\right)\,T_{k,\ell}
=0.
\end{equation}
\end{prop}
\noindent
Proof is similar to that of Proposition \ref{slprop}.

Let us now calculate the dimensions of the algebras ${\cI^k_{\l,\m}}^{\rm loc}(S^1)$.
\begin{thm}
\label{dim}
The  dimensions of the algebras ${\cI^k_{\l,\m}}^{\rm loc}(S^1)$ are given
in the following table
$$
\begin{array}{|l|c|c|c|c|c|c|}
\hline
k & 0 & 1&2&3&4&\geq 5\\[4pt]\hline
(\l,\m)\,\hbox{generic} &1&2&2&1&1&1\\[4pt]\hline
\l=0,\,\hbox{or}\,\m=1,\,\hbox{generic} &1&2&3&3&2&2\\[4pt]\hline
\l + \mu =1,\,\hbox{generic} &1&2&2&2&2&2\\[4pt]\hline
(3\l+1)(3\m-4)=-1,\,\hbox{or}\,\m-\l=2, \,\hbox{generic}  &1&2&2&2&1&1\\[4pt]\hline
(\l,\m)=(-\frac{1}{4},1),\,(-2,1) ,\,(0,\frac{5}{4}),\,(0,3)&1&2&3&3&3&2\\[4pt]\hline
(\l,\m)=(0,0),\,(1,1) &1&2&3&3&3&3\\[4pt]\hline
(\l,\m)=(-\frac{2}{3},\frac{5}{3})&1&2&2&3&3&2\\[4pt]\hline
(\l,\m)=(-\frac{1}{2},\frac{3}{2})&1&2&3&3&2&2\\[4pt]\hline
(\l,\m)=(0,1)&1&3&4&5&5&5\\[4pt]\hline
\end{array}
$$
\end{thm}

\begin{proof}
We will use the following result (which is similar to that of \cite{LMT}).

\begin{prop}
\label{propDet}
Every $\Diff(S^1)$-invariant differential operator $T$ on $\cD^k_{\l,\m}(S^1)$ with
$k\geq3$ is completely determined by its restriction to the subspace of third-order
operators
$\left.T\right|_{\cD^3_{\l,\m}(S^1)}$.
\end{prop}

\begin{proof}
Assume that $k\geq 4$ and $\left.T\right|_{\cD^3_{\l,\m}(S^1)}=0$ that is, $T_{r,\ell}=0$
for all $\ell$ and all $r\leq 3$. Then the system of equations
(\ref{relat},\ref{LastEq}) readily leads to $T_{r,\ell}=0$ for all $(r,\ell)$.
\end{proof}

The end of the proof of Theorem \ref{dim} is as follows.

We solve the system (\ref{relat}), (\ref{LastEq}) explicitly for $k\leq5$ (we omit here the
tedious computations) and obtain the result in this case.

It follows from Proposition~\ref{propDet} that the dimension of the algebras 
${\cI^k_{\l,\m}}^{\rm loc}(S^1)$ of local symmetries with
$k\geq 3$ can only decrease as $k$ becomes
$k+1$. 
On the other hand, we have already constructed a set of generators of the algebra 
$\cI^k_{\l,\m}(S^1)$ that gives a lower bound for the dimension.
We thus conclude, by
Proposition~\ref{propDet}, that the constructed generators span the algebras
$\cI^k_{\l,\m}(S^1)$ for all $k$.

Theorem \ref{dim} is proved.
\end{proof}

\subsection{Non-local operators}

Consider now the non-local case. We already constructed an example of a non-local linear
map (\ref{Endom}) commuting with the $\Vect(S^1)$-action, see formula (\ref{leL}). Let us
show that there are no other such maps.

Let $T$ be a non-local linear map (\ref{Endom}) commuting with the $\Vect(S^1)$-action.
Assume that $A\in\cD^k_{\l,\m}(S^1)$ vanishes on an open subset $U\subset{}S^1$, 
but $T(A)\in\cD^k_{\l,\m}(S^1)$ does not
vanish on $U$. Let $X$ be a vector field with support in $U$, 
then $\cL^{\l,\m}_X(A)=0$.
Since $T$ satisfies the relation of equivariance
$$
\cL^{\l,\m}_X(T(A))=T(\cL^{\l,\m}_X(A)),
$$
one gets
$$
\cL^{\l,\m}_X\left(T(A)\right)=0.
$$
Consider the restriction $T(A)|_U$, this is an element of $\cD^k_{\l,\m}(U)$. We just proved
that $T(A)|_U$ is a $\Vect(S^1)$- and thus a $\Diff(S^1)$-invariant differential operator
$\cF_\l(U)\to\cF_\m(U)$.

The classification of such invariant differential operators (on any manifold) is well known
(see, e.g., \cite{Fuc,Gro} and \cite{LLS} for proofs); the answer is as follows. There exists
a unique non-trivial invariant differential operator, namely the de Rham differential.
Therefore
$T(A)|_U$ is proportional to one of the operators: 
\begin{itemize}
\item
$\Id\in\cD^k_{\l,\l}(U)$, so that $\m=\l$, or
\item
$d\in\cD^k_{0,1}(U)$, so that $(\l,\m)=(0,1)$.
\end{itemize}
In each case, one gets an invariant linear functional, namely, the operator $T$ is of the
form
\begin{itemize}
\item
$T=t\,\Id$, where $t:\cD^k_{\l,\l}(U)\to\bbR,$
or
\item
$T=t\,d$, where
$t:\cD^k_{0,1}(U)\to\bbR$,
\end{itemize}
respectively. The linear functional $t$ has to be $\Diff(S^1)$-invariant.

The space of symbols corresponding to the above modules are
$$
\cF_0\oplus\cdots\oplus\cF_{-k}
\qquad
\hbox{and}
\qquad
\cF_1\oplus\cdots\oplus\cF_{1-k},
$$
respectively. Projecting the functional $t$ to these modules of symbols one obtains a
linear functional which is again $\Diff(S^1)$-invariant. Moreover,
this functional is non-zero if and only if
$t$ is non-zero.

\begin{lem}
\label{FuncLem}
There exists a unique (up to a multiplicative constant) non-trivial invariant functional 
$\cF_\l(S^1)\to\bbR$ if and only if $\l=1$. This functional is the integral of 1-forms
$$
\int:\cF_1(S^1)\to\bbC
$$
\end{lem}
\begin{proof}
The statement follows from the fact that any module of tensor fields on a manifold $M$ is
irreducuble except the modules of differential forms $\Omega^k(M)$, see \cite{Rud}.
However, in the one-dimensional case, one can give simple direct arguments.

Let $\tau:\cF_\l(S^1)\to\bbR$ be a $\Diff(S^1)$-invariant linear functional. Then
$\ker\tau\subset\cF_\l(S^1)$ is a submodule. For every vector field $X=X(x)\frac{d}{dx}$ and
every $\l$-density $\varphi=\phi(x)(dx)^\l$, one has 
$$
\tau(L_X\varphi)=0.
$$
Consider the function $\psi(x)$ defined by the expression
$$
\phi(x)=X(x)\phi'(x)+\l\,X'(x)\phi(x).
$$
It is easy to check that, in the case $\l\neq1$, for every $\psi(x)$ there exist $X(x)$
and $\phi(x)$ such that the above equation is satisfied. If $\l=1$ then $\psi(x)$ has to
satisfy the property $\int\psi(x)\,dx=0$.
\end{proof}

The lemma implies that the functional $t$ can be non-zero only in the second case and has to
be proportional to (\ref{nonloc}). We conclude that every non-local $\Diff(S^1)$-invariant
operator $T$ has to be proportional to the operator $L$ given by (\ref{leL}).

Theorems~\ref{Dim1Thm1}-\ref{Dim1Thm5} are proved.

\section{Conclusion and outlook}
\label{CotSec}

The classification theorems of Section \ref{MainResSec} provide a number of particular
examples of modules of differential operators. Some of these modules are known (however,
the precise values of parameters $(\l,\m)$ are often implicit), other ones are new.

\subsection{Known modules of differential operators}

There are particular examples of modules of differential operators on $S^1$ that have been
known for a long time. Some of these modules appear in our
classification. Let us briefly mention here some interesting cases.

The family of modules $\cD^k_{\l,\m}(S^1)$ with $(\l,\m)$ satisfying the condition $\l+\m=1$
is, of course, the best known class. This is the only case when one can speak of conjugation
and split a given differential operator into the sum of a symmetric and a skew-symmetric
operators.

The module $\cD^2_{-\half,\frac{3}{2}}(S^1)$
is a well-known module of second-order differential operators. In contains a submodule
of Sturm-Liouville operators
$$
A=\frac{d^2}{dx^2}+a(x)
$$
which is related to the Virasoro algebra (see \cite{KIR}).
This module also has a very interesting geometric meaning: it is isomorphic to the space
of projective structures on $S^1$ (see \cite{KIR} and also \cite{OT}). 

The value $\l=-\half$ is also related to the Lie \textit{superalgebras} of Neveu-Schwarz and
Ramond (see \cite{KIR2}). More precisely, the odd parts of these superalgebras consist of
$-\half$-densities.

The above module is a particular case of the following series of modules.

The modules $\cD^k_{\frac{1-k}{2},\frac{1+k}{2}}(S^1)$, see formula
(\ref{WilMod}), also have interesting geometric and algebraic meaning. 
These modules have already been studied in \cite{WIL}, it turns out that they are related to
the space of curves in the projective space $\bbP^{k-1}$ (see, e.g.,
\cite{OT}). These modules are also related to so-called Adler-Gelfand-Dickey bracket (see,
e.g., \cite{OO} and references therein).

\subsection{New examples of modules of differential operators}

Some of the particular modules $\cD^k_{\l,\m}(S^1)$ provided by our classification theorems
seem to be new.

The modules (\ref{hk}) can be characterized as the modules that are ``abnormally close'' to
the corresponding $\Diff(S^1)$-modules of symbols. More precisely, in this case, the
quotient-module
$$
\cD^k_{\l,\m}(S^1)/\cD^{k-3}_{\l,\m}(S^1)\cong\cF_{\d-k}\oplus\cF_{\d-k+1}\oplus\cF_{\d-k+2}
$$
where $\d=\m-\l$.
The operator (\ref{maphk}) has been known to the classics in some particular cases (cf.
\cite{WIL}).

An interesting particular case that belongs to the above family is the module
$\cD^3_{-\frac{2}{3},\frac{5}{3}}(S^1)$. It can be characterized by three conditions:
$k=3$ together with $\l+\m=1$ and (\ref{hk}). This module is related to the Grozman operator
(\ref{Gro}).

The module $\cD^4_{-\frac{2}{3},\frac{5}{3}}(S^1)$ is also related to the Grozman operator.
We strongly believe that this exceptional module has an interesting geometric and algebraic
meaning.

Other examples of exceptional modules of Theorems \ref{Dim1Thm2} and \ref{Dim1Thm3}, such as
$\cD^4_{-\frac{1}{4},1}(S^1)$,
$\cD^4_{-2,1}(S^1)$ or $\cD^3_{-1,1}(S^1)$, etc. are even more mysterious.

\subsection{Results in the multi-dimensional case}

For the sake of completeness, let us mention the results in the multi-dimensional case.
Let $M$ be a smooth manifold, $\dim M\geq2$; the classification of invariant
operators on $\cD^k_{\l,\m}(M)$ was obtained in \cite{MAT}.
The algebra of symmetries $\cI^k_{\l,\m}(M)$ does not depend on the topology of $M$. This
algebra is trivial for all $(\l,\m)$, except for the cases
$\l+\m=1$ and $\l=0$ or $\m=1$. The generators are $C,P_0$ and $\Id$.

\section{Appendix: generators of the algebras $\fa$, $\fb$ and $\bbR^n$}
\label{Appen}

Let us introduce the generators of the matrix algebras $\fa$, $\fb$ and $\bbR^n$ that are
useful in order to establish the isomorphisms with the algebras of symmetry
$\cI^k_{\l,\m}(S^1)$. 

In the case of the algebra $\fa$,
we take the matrices
$$
\tilde{a}=
\left(\begin{array}{cc}1 & 0\\ 0 & 1\end{array}\right),
\quad
\tilde{b}=
\left(\begin{array}{cc}0 & 0\\ 1 & 0\end{array}\right).
$$
The relations between the generators are:
$$
\tilde{a}^2=\tilde{a},
\qquad
\tilde{a}\tilde{b}=\tilde{b}\tilde{a}=\tilde{b},
\qquad
\tilde{b}^2=0.
$$

For the algebra $\fb$, we consider
$$
\scriptstyle
{\textstyle\bar{a}}
=\left(\begin{array}{cccc}
1 & 0 & 0 & 0\\
0 & 1 & 0 & 0\\
0 & 0 & 0 & 0\\
0 & 0 & 0 & 0
\end{array}\right),
\quad
{\textstyle\bar{b}}=\left(\begin{array}{cccc}
0 & 0 & 0 & 0\\
0 & 0 & 0 & 0\\
0 & 0 & 1 & 0\\
0 & 0 & 0 & 1
\end{array}\right),
\quad
{\textstyle\bar{c}}=\left(\begin{array}{cccc}
0 & 0 & 0 & 0\\
0 & 0 & 0 & 0\\
0 & 1 & 0 & 0\\
0 & 0 & 0 & 0
\end{array}\right),
\quad
{\textstyle\bar{d}}=\left(\begin{array}{cccc}
0 & 0 & 0 & 1\\
0 & 0 & 0 & 0\\
0 & 0 & 0 & 0\\
0 & 0 & 0 & 0
\end{array}\right).
$$
The multiplication table for this algebra is
$$
\begin{array}{c||c|c|c|c}
 & \bar{a} & \bar{b} & \bar{c} & \bar{d} \\[4pt]\hline\hline
\bar{a} & \bar{a} & 0 & 0 & \bar{d} \\[4pt]\hline
\bar{b} & 0 & \bar{b} & \bar{c} & 0 \\[4pt]\hline
\bar{c} & \bar{c} & 0 & 0 & 0 \\[4pt]\hline
\bar{d} & 0 & \bar{d} & 0 & 0 
\end{array}
$$

In the case of the algebra $\bbR^n$, one can choose the generators
$1,\bar{a}_1,\ldots\bar{a}_{n-1}$ with relations:
$$
1\,\bar{a}_i=\bar{a}_i\,1=\bar{a}_i,
\qquad
\bar{a}_i\bar{a}_j=\d_{ij}.
$$

\textbf{Acknowledgments:}
We are pleased to thank D. Leites for his help.
We are grateful to C.~Duval, A.~El Gradechi, 
P.~Lecomte and S. Parmentier for
enlightening discussions and their interest in this work.
The third author is grateful to Swiss National Science Foundation
for its support and to A. Alekseev for hospitality.

\vskip 1cm



\begin{thebibliography}{99}

\bibitem{Adl}
Adler M,
On a trace functional for formal pseudo differential operators and 
the symplectic structure of the Korteweg-de Vries type equations,
{\it Invent. Math.} {\bf 50} (1978/79), 219--248.

\bibitem{BO} 
Bouarroudj S, Ovsienko V,
Three cocycles on
$\Diff(S^1)$ generalizing the Schwarzian derivative,
{\it Internat. Math. Res. Notices} {\ bf 1} (1998), 25--39.

\bibitem{CAR} 
Cartan E, 
Le\c cons sur la th\'eorie des espaces
\`a connexion projective, Gauthier -Villars, Paris, 1937.

\bibitem{DO} 
Duval C, Ovsienko V,
Space of second order linear differential operators as a module over the 
Lie algebra of vector fields, 
{\it Adv. Math.} 
{\bf 132:2} (1997), 316--333.

\bibitem{FF} 
Feigin B L, Fuchs D B, 
Homology of the Lie algebra of vector
fields on the line, 
{\it Func. Anal. Appl.} {\bf 14} (1980),
201--212.

\bibitem{FF1} 
Feigin B L, Fuchs D B, 
Skew-symmetric invariant
differential operators on the line and Verma modules over the Virasoro algebra. (Russian) 
{\it Func. Anal. Appl} {\bf 16:2}  (1982), 47--63.

\bibitem{Fuc} 
Fuchs D B, 
Cohomology of infinite-dimensional Lie
algebras, Consultants Bureau, New York, 1987.

\bibitem{GAR} 
Gargoubi H,
Sur la g\'eom\'etrie de l'espace des op\'erateurs diff\'erentiels
lin\'eaires sur $\bf R$,
{\it Bull. Soc. Roy. Sci. Li\`{e}ge} {\bf 69:1} (2000), 21--47. 

\bibitem{GO} 
Gargoubi H, Ovsienko V, 
Space of linear differential
operators on the real line as a module over the  Lie algebra of vector fields,
{\it Int. Math. Res. Not.} {\bf 5} (1996), 235--251. 

\bibitem{GO1} 
Gargoubi H, Ovsienko V, 
Modules of differential operators on the real line,
{\it Funct. Anal. Appl.} {\bf 35:1} (2001), 13--18.

\bibitem{Gro} 
Grozman P Ya,
Classification of bilinear invariant operators over tensor fields,
{\it Functional Anal. Appl.} {\bf 14:2} (1980), 127--128.

\bibitem{GLS}
Grozman P, Leites D, Shchepochkina I, 
Invariant operators on supermanifolds and standard models,
in: Multiple facets of quantization and supersymmetry,  
508--555, M. Olshanetski, A. Vainstein (eds.), 
World Sci. Publishing, 2002, math.RT/0202193.

\bibitem{KIR}  
Kirillov A A, 
Infinite dimensional Lie
groups~: their orbits, invariants and representations. The geometry of
moments,  Lect. Notes in Math., {\bf 970}
Springer-Verlag (1982), 101--123.


\bibitem{KIR1}  
Kirillov A A, 
Invariant operators over 
geometric quantities (Russian), 
in: Current Problems in Mathematics, {\bf 16}, 3--29, 
Akad. Nauk SSSR, VINITI, Moscow, 1980;
[English translation: J. Sov. Math {\bf 18:1} (1982), 1--21].

\bibitem{KIR2}  
Kirillov A A, 
The orbits of the group of diffeomorphisms of the
circle, and local Lie superalgebras,  
{\it Func. Anal. Appl.}  {\bf 15:2}  (1981), 75--76.

\bibitem{KMS} 
Kol\'ar I, Michor P, Slov\'ak J, 
Natural operations in differential geometry. 
Springer-Verlag, Berlin, 1993.

\bibitem{LLS} 
Lebedev A, Leites D, Shereshevskii I,
Lie superalgebra structures in cohomology spaces of Lie algebras with
coefficients in the adjoint representation,
{\it math.KT/0404139}.

\bibitem{LMT} 
Lecomte P B A, Mathonet P, Tousset E,
Comparison of some modules of the Lie algebra of vector fields,
{\it Indag. Mathem., N.S.} {\bf 7:4} (1996), 461--471.

\bibitem{LO} 
Lecomte P B A, Ovsienko V,
Projectively invariant symbol calculus,
{\it Lett. Math. Phys.} {\bf 49:3} (1999), 173--196.

\bibitem{LO1} 
Lecomte P B A, Ovsienko V,
Cohomology of the vector fields Lie algebra and modules of
differential operators on a smooth
manifold, 
{\it Compos. Math.} {\bf 124:1} (2000), 95--110.

\bibitem{MP} 
Martin C, Piard A, 
Classification of the indecomposable bounded admissible
modules over the Virasoro Lie algebra with weightspaces of 
dimension not exceeding two, 
{\it Comm. Math. Phys.} {\bf 150:3}
(1992), 465--493.

\bibitem{MAT} 
Mathonet P,
Intertwining operators between some spaces of differential
operators on a manifold, 
{\it Comm. Algebra} {\bf 27:2}
(1999), 755--776.

\bibitem{MAT1} 
Mathonet P,
Geometric quantities associated to differential
operators, 
{\it Comm. Algebra} {\bf 28:2} (2000),
699--718.

\bibitem{OO} 
Ovsienko O, Ovsienko V, 
Lie derivatives of order $n$ on the line. Tensor meaning
of the Gelfand-Dikii bracket, 
{\it Adv. Soviet Math.} {\bf 2} (1991), 221--231.

\bibitem{OT} 
Ovsienko V, Tabachnikov S,
Projective differential geometry old and new: from the Schwarzian derivative to cohomology
of diffeomorphism groups, Cambridge University Press, 2004.

\bibitem{Pee} 
Peetre J,
Une caract\'erisation abstraite des op\'erateurs
diff\'erentiels,
{\it Math. Scand.} {\bf 7} (1959), 211--218 and {\bf 8} (1960),
116--120.

\bibitem{Rud} 
Rudakov A N,
Irreducible representations of
infinite-dimensional Lie algebras of Cartan type,
{\it Izv. Akad. Nauk SSSR Ser. Mat.} {\bf 38} (1974), 835--866.

\bibitem{SEG} 
Segal G B,
Unitary representations of some
infinite dimensional groups, 
{\it Comm. Math. Phys.} {\bf 80:3}
(1981), 301--342.

\bibitem{Ter} 
Terng C L,
Natural vector bundles and natural differential operators, 
{\it Amer. J. Math.}
{\bf 100:4} (1978), 775--828.

\bibitem{Veb}
Veblen O,
Differential invariants and geometry,
Atti del Congresso Internazionale dei Matematici,
Bologna, 1928.

\bibitem{Wey}
Weyl H,
{\sl The Classical Groups}, 
Princeton University Press, 1946.


\bibitem{WIL} 
Wilczynski, E J
Projective differential geometry
of curves and ruled surfaces, Leipzig - Teubner, 1906.

\end{thebibliography}
\end{document}